\documentclass[review,3p, 12pt]{elsarticle}

\usepackage{amsmath,amsmath,mathrsfs}
\usepackage{color,fancyhdr,epsf,psfrag,lineno,hyperref,txfonts,subfigure}
\usepackage{multirow}
\usepackage{graphicx}
\usepackage{mhchem}
\graphicspath{ {figs/} }
\usepackage{bm,ulem}

\definecolor{R1}{rgb}{0.0,0.0,0.0} 
\definecolor{R3}{rgb}{0.0,0.0,0.0} 
\definecolor{R2}{rgb}{0.0,0,0.0} 

\journal{Aerospace Science and Technology}

\biboptions{sort&compress}

\begin{document}

\begin{frontmatter}

\title {Direct numerical simulation of inflow boundary-layer turbulence effects on cavity flame stabilisation in a model scramjet combustor}  

\author[a]{Minqi Lin}
\author[b]{Jian Fang}
\author[c]{Xi Deng}
\author[b]{Xiaojun Gu}
\author[a,d]{Zhi X. Chen\corref{cor1}}
    \cortext[cor1]{Corresponding author.}
	\ead{chenzhi@pku.edu.cn}

\address[a]{State Key Laboratory for Turbulence and Complex Systems, College of Engineering, Peking University, Beijing 100871, P.R. China}
\address[b]{Scientific Computing Department, Science and Technology Facilities Council (STFC), Daresbury Laboratory, Warrington WA4 4AD, United Kingdom}
\address[c]{Department of Engineering, University of Cambridge, Cambridge CB2 1TN, United Kingdom}
\address[d]{AI for Science Institute (AISI), Beijing, 100080, P.R. China}

\begin{abstract}
Supersonic lean premixed hydrogen/air combustion stabilised by a cavity-flame holder within a model scramjet, characterized by a Mach 1.5 inflow at 1000~K and 50~kPa, is investigated via direct numerical simulation. By separately implementing wall-bounded turbulent and laminar inlet conditions, this work analysis various physical processes of flame stabilization and turbulence-flame interactions to study the influence of inflow boundary layer conditions.
Findings indicate that combustion occurred within the cavity shear layer in both cases and propagated downstream along the lower wall. Also, the server impingement at the rear wall in the case with laminar inflow leads to greater cavity resistance. Furthermore, the studies on gas exchange and transport process indicates that with laminar inflow the entered gas accumulates in the back part of the cavity via the intensive mass exchange process and weaker interaction between the primary and secondary vortices. Flame stretch and thickness are further investigated to shed light into turbulence-flame interaction in supersonic flows. Findings indicates that the case with inflow wall-bounded turbulence show similar behaviours compared to previous studies, whereas the observed phenomena in front part of cavity shear layer are differ due to the presence of roll-up vortices in the case with laminar inflow. Overall, the influence of tangential strain rate and curvature are consistent with the preceding results and the evolution of flame thickness is caused by the combined effect of the two factors in both cases.

\vspace{8 pt}
\noindent
\textbf{Highlights}

\begin{itemize}

  \item{Direct numerical simulation conducted on a laboratory-scale model scramjet combustor under supersonic mode is a pioneering attempt in the community 
  and the importance of high-fidelity data further underscores the value of this study.}
    
  \item{The analysis of the flame-vortex interaction based on the highly-resolved simulation have the potential for making inspiring contributions to the society.} 
  

  \item{ Conclusions can be drawn through meticulously depicted graphs of detailed physical structures without extensive background information. }

\end{itemize}





\end{abstract}

\vspace{12 pt}

\end{frontmatter}

\section{Introduction}\label{introduction}
In air-breathing scramjets, the presence of relatively high Mach numbers within the combustor results in a limited residence time for the fuel-oxidation mixture to ignite and combustion propagation~\cite{urzay2018supersonic}. Therefore, flame stabilization emerges as a critical and complex challenge that necessitates enhancement. Numerous fundamental research studies have been conducted both experimentally and numerically (see Ref.~\cite{liu2020review} for a detailed review).
An effective method of holding the flame in the combustion chamber is the use of a wall cavity, which was adopted by Roudakov et al.~\cite{roudakov1993flight} and subsequently confirmed by experiments to improve the efficiency of the combustion of hydrocarbons in the supersonic flow~\cite{vinogradov1995experimental,ortwerth1996experimental,owens1998flame}. 
Numerical simulations have been performed across various cavity configurations to uncover the flame stabilization mechanisms and to validate combustion performance.
Previous studies have shown that open cavities are preferred due to their lower drag penalty compared to closed cavities~\cite{vortex,BARNES201524}. 
Recent studies have proposed alternative cavity geometries, such as angled rear walls~\cite{sun2020unsteady, ben2001cavity}, to further improve flame stability. However, it should be noted that the incorporation of angled rear walls may influence the drag penalty of the cavity. Ben-Yakar et al.~\cite{ben2001cavity} indicated that there exists a critical angle of the back wall, estimated to be between 16 and 45 degrees, at which the drag penalties of a cavity are minimised.
Therefore,  it is inferred an open cavity with an appropriately slanted trailing edge is an effective method for flame-holding in supersonic combustors.

To exolore the underlying physical mechanism, numerical studies have been conducted. 
Davis et al.~\cite{Davis1997} performed Reynolds-averaged Navier-Stokes (RANS) simulations which demonstrated a strong concordance with experiments for flame stabilization in recessed cavities.
Gruber et al.~\cite{vortex} and Kim et al.~\cite{KIM2004271} tested various angles of the aft wall and discussed their performances. Several studies~\cite{liu2006simulations,Yang2010} adopted the RANS approach to simulate the flame under different flow conditions, such as with and without air throttling, and different injection locations based on combustors designed by the U.S. Air Force. 
Studies have shown that the RANS approach is accurate in simulating mean flow in the combustion chamber, but the use of turbulence models can still produce large uncertainties~\cite{keistler2005turbulence}. Therefore, Large-eddy simulation (LES) has been used in studies~\cite{sun2009experimental,sun2011flame,ghodke2011large} to investigate the flame stabilisation mechanism and to analyse the unsteady flame characteristics. In recent studies, Ruan et al.~\cite{Ruan2020238} recently applied LES to learn the preferential combustion modes in cavity-based scramjet and results of flow fields shown great agreement with experiments. Using a hybrid RANS/LES method, Wang et al.~\cite{wang2013combustion} investigated various effects on the supersonic combustion characteristics, such as the jet-cavity interaction through mass exchange process~\cite{wang2014large} and effect of different angled aft walls~\cite{wang2014experimental}.

In recent studies, highly resolved simulations have been conducted to elucidate the turbulent flame-vortex interaction. Goodwin et al.~\cite{goodwin2020effect} adopted the DG method~\cite{johnson2020conservative} to simulate premixed ethylene-air combustion under subsonic turbulent inflow conditions and obtained highly resolved results. Rising et al.~\cite{rising2022numerical} utilized the same method and further examined flame-vortex interaction under different geometry. Recently, Sitaraman et al.~\cite{SITARAMAN2021111531} compared two cases with different injection positions on the bottom wall of the cavity. In their study, second-order numerical schemes incorporating adaptive mesh refinement were employed to achieve higher numerical accuracy, particularly in regions adjacent to shock waves and within the cavity.
Recent advancements in high-performance computing (HPC) technology have rendered direct numerical simulation (DNS) a viable methodology for simulating supersonic reactive flows, applicable to both simplified geometries~\cite{mahle2007turbulence,ferrer2017compressibility,o2014subgrid} and realistic configurations of scramjet flameholders~\cite{fang2023direct,linICDERS2023,SITARAMAN2021111531,goodwin2020effect,rising2022numerical}.
Aditya et al.\cite{aditya2019dns} used the eighth-order S3D code~\cite{JHChen2009} to perform DNS of a turbulent premixed ethylene-air flame over a backward-facing step, albeit with a simpler geometry. Rauch et al.\cite{rauch2018dns} employed the same code to conduct two-dimensional and three-dimensinal DNS investigations of premixed turbulent flames with rectangular and linear ramp cavities, demonstrating that the linear ramp cavity facilitates better fuel entry into the cavity. However, both studies mentioned above were conducted under subsonic inflow conditions.

In this work, we present a high-resolution numerical simulation
study on the supersonic lean premixed H$_2$-air combustion in a model scramjet combustor. The simulated configuration has a free-stream Mach number of 1.5 at temperature of 1000~K and pressure of 0.5~bar. The geometry of wall cavity mimics the size of the dual-mode, direct-connect combustor at the University of Virginia Supersonic Combustion Facility~\cite{cutler2018coherent}. This work is a follow-up of our earlier study on the non-reactive flows for the same configuration~\cite{fang2023direct}, with the aims to shed light into the structure of cavity stabilised flame. 
In a previous experimental study, Liu et al.~\cite{liu2017influences} investigated the effect of free-stream turbulence on the supersonic flame dynamics and demonstrated that the intensity of the incoming turbulence significantly influences flame burning behaviours and stability.
However, due to the limitation of measurement techniques under the extreme conditions of scramjets, the detailed physical mechanism driving these flame behaviours remain unclear. To the best knowledge of the authors, the influence of supersonic inflow wall-bounded turbulence conditions have not yet been studied via scale-resolved simulations in the past literature. To this end, a particular focus of this paper is given on the several inflow conditions and their effects on the cavity-stabilised flame characteristics are investigated in detail. 

The remainder of this paper is organised as follows. Computational methodology and test case details are described in Section~\ref{sec:Simulation}. In Section~\ref{sec:result}, simulation results focused on the area near the cavity are presented and discussed. Finally, the conclusions are summarised in Section~\ref{sec:con}.

\section{Simulation Details}
\label{sec:Simulation}

\subsection{Numerical method}
The three-dimensional compressible Navier-Stokes (N-S) equations are solved in this simulation. Considering multi-species chemical reaction, the governing equations are written as
\begin{equation}
\frac{\partial \rho}{\partial t} +
\frac{\partial \rho u_i}{\partial x_i} = 0 \:,
\label{eq:1}
\end{equation}
\begin{equation}
\frac{\partial \rho u_i}{\partial t} +
\frac{\partial \rho u_iu_j}{\partial x_j} = 
-\frac{\partial p}{\partial x_i}
+\frac{\partial \sigma_{ij}}{\partial x_j} \:,
\end{equation}
\begin{equation}
\frac{\partial \rho e}{\partial t} +
\frac{\partial (\rho e+p)u_i}{\partial x_i} = 
-\frac{\partial q_i}{\partial x_i}
+\frac{\partial \sigma _{ij}u_i}{\partial x_j} \:,
\end{equation}
\begin{equation}
\frac{\partial \rho Y_k}{\partial t} +
\frac{\partial \rho Y_k(u_i+V_{k,i})}{\partial x_i} = 
\dot{\omega}_k  \:
\label{eq:4}
\end{equation}
where $\rho$ is the density, $u_i$ is the velocity component in $i$ direction, $p$ is the pressure, $e$ is the total energy per unit mass. $Y_k$, $D_k$ and $\dot{\omega}_k$ represent the mass fraction, coefficient of mass diffusion, and mass-based chemical source term of the $k$-th species, respectively. The stress tensor and heat flux vector are given by 
\begin{equation}
\sigma_{ij} = \mu (
\frac{\partial u_i}{\partial x_j} + \frac{\partial u_j}{\partial x_i}
-\frac{2}{3} \delta_{ij} \frac{\partial u_k}{\partial x_k} ) \: ,
\end{equation}
\begin{equation}
q_i = - \lambda \frac{\partial T}{\partial x_i} + \rho \sum_{k=1}^{N}h_kY_kV_{k,i} \: ,
\end{equation}
where $\mu$ denotes the fluid viscosity, $\lambda$ is the coefficient of the heat conductivity, $h_k$ is the enthalpy of the $k$-th species and $V_{k,i}$ is the diffusion velocity of the specie $k$ in the $i$ direction. The diffusion velocity is computed as
\begin{equation}
 V_{k,i} =  - \frac{1}{X_k} D_k \frac{\partial X_k}{\partial x_i} + \sum\limits_{k}\frac{Y_k}{X_k} D_k \frac{\partial X_k}{\partial x_i} \: ,
\end{equation}
\begin{equation}
    D_k = \frac{1-Y_k}{\sum_{n\neq k} X_n/D_{nk}} \: ,
\end{equation}
where $X_k$ is the mole fraction of the $k$-th species, $D_k$  is the mixture-averaged diffusion coefficient and $D_{nk}$ is the binary diffusion coefficients. Thermal diffusion (Soret effect) is neglected in this work. For the chemical system with $N$ species and $M$ reactions:
\begin{equation}
\sum_{k=1}^{N} {\nu}'_{kj} \mathcal{M}_k\rightleftharpoons \sum_{k=1}^{N} {\nu}''_{kj} \mathcal{M}_k \ \ \ \mathrm{for} \ \ \ j=1,\dots,M\:,
\end{equation}
where $\mathcal{M}_k$ represent the specie $k$, ${\nu}'_{kj}$ and ${\nu}''_{kj}$ represent the molar stoichiometric coefficients of species $k$ in reaction $j$. Then the chemical source term of the $k$-th species $\dot{\omega}_k$ can be computed as
\begin{equation}
\dot{\omega}_k = W_k\sum_{j=1}^{M} (\nu''_{kj}-\nu'_{kj})\left \{ K_{fj} {\textstyle \prod_{\alpha=1}^{N}}(\frac{\rho Y_\alpha}{W_\alpha})^{\nu '_{\alpha j}} -K_{rj} {\textstyle \prod_{\alpha=1}^{N}} (\frac{\rho Y_\alpha}{W_\alpha})^{\nu ''_{\alpha j}} \right \} \:,
\end{equation}
where $K_{fj}$, $K_{rj}$ represent the forward and reverse rates of reaction $j$ calculated using the Arrhenius law, respectively. $W_k$ is the molecular weight for the $k$-th species.

The finite difference method with compact stencils is adopted to solve the transport equations. An in-house high-order computational fluid dynamics code ASTR is used. This code has been applied to a series of numerical simulation for high-speed flows and interaction between shock wave and turbulent boundary layer~\cite{fang2014,fang2015direct,fang2017investigation,fang2020turbulence,fang2023direct}, and is coupled with the Cantera Fortran interface~\cite{cantera} to calculate chemical kinetics, thermal and transport properties. The 9 species, 12 reactions hydrogen-air combustion mechanism~\cite{boivinExplicitReducedMechanism2011} is used in this work. The convection terms of Eq.~(\ref{eq:1}) $\sim$ (\ref{eq:4}) are approximated by a seventh-order low-dissipative monotonicity-preserving scheme~\cite{fang2013optimized} in order to preserve accuracy near shock-waves. As for the diffusion terms, a sixth-order central scheme is adopted and the mixture-averaged approximation is applied as the molecular diffusion model. The temporal integration is conducted by a low-storage three-step third-order Runge-Kutta method. Further details regarding the numerical implementation and code validation can be found in Refs.~\cite{Xu2023TGVflame,fang2023direct}.

\subsection{Configuration description}
\begin{figure}[t]
    \centering\centering\ifx\mycmd\undefined
    \includegraphics[width=0.8\textwidth]{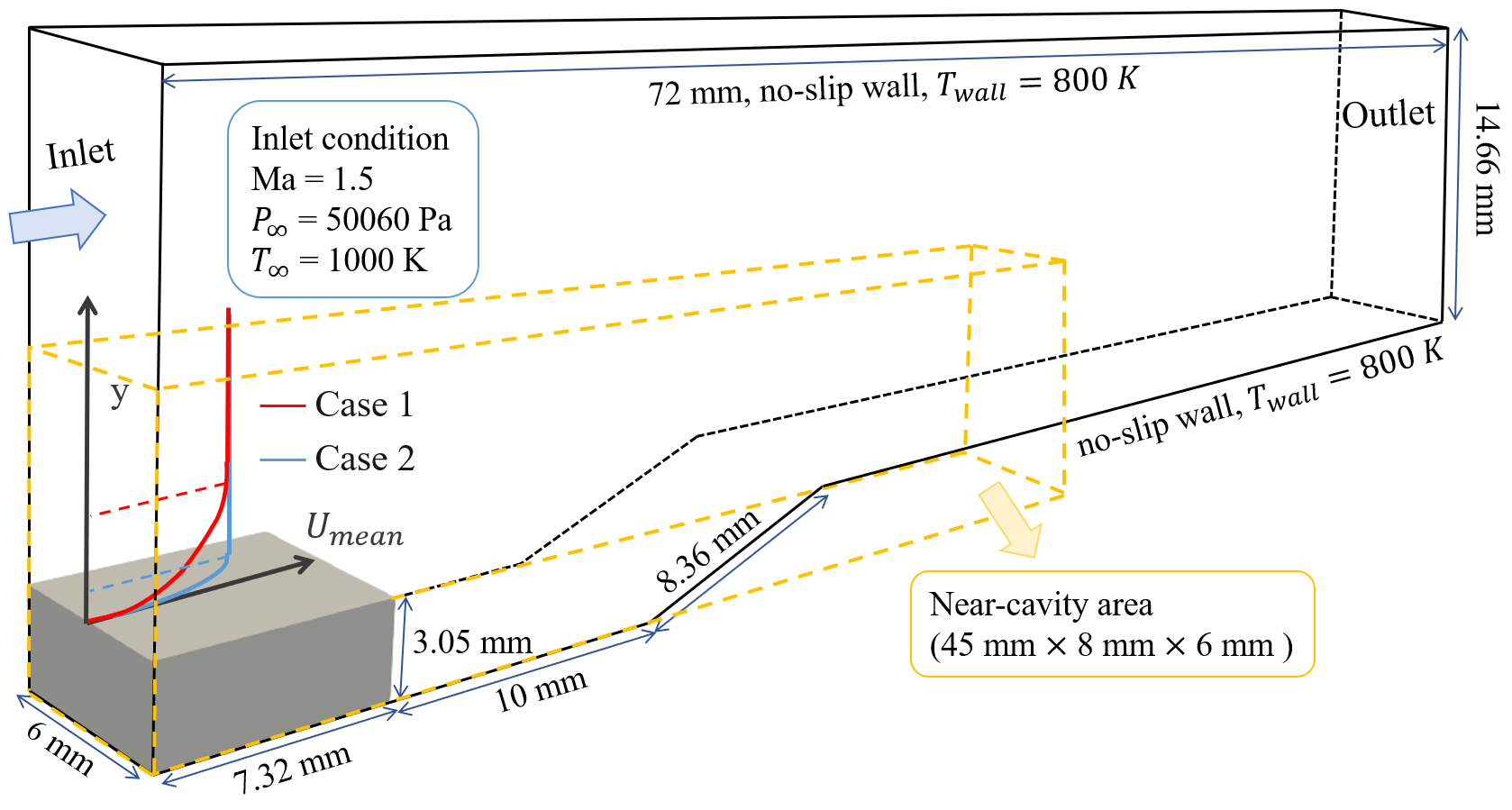}
    \fi
    \caption{Schematic of computational domain with inflow conditions of two cases.} 
    \label{fig1:geom}
\end{figure}

The computational domain involved a quasi-3D planar channel with a cavity embedded in the lower wall. Specific geometry sizes marked in Fig.~\ref{fig1:geom} are set refer to the dual-mode, direct-connect combustor at the University of Virginia Supersonic Combustion Facility. 

An immersed boundary method, presented by Vanna et al.~\cite{de2020sharp} with some modifications, is used for the upstream wall and front wall of the cavity. Because of the limited impact of different wall thermal condition observed in study of Ruan et al.~\cite{Ruan2020238}, non-slip wall boundary with fixed wall temperature of 800~K is adapted to the upper and lower wall to avoid excessive temperature. The periodic condition is applied in the spanwise direction and a non-reflective outflow condition is applied at the outlet plane.

\begin{figure}[t]
    \centering\centering\ifx\mycmd\undefined
    \includegraphics[width=0.9\textwidth]{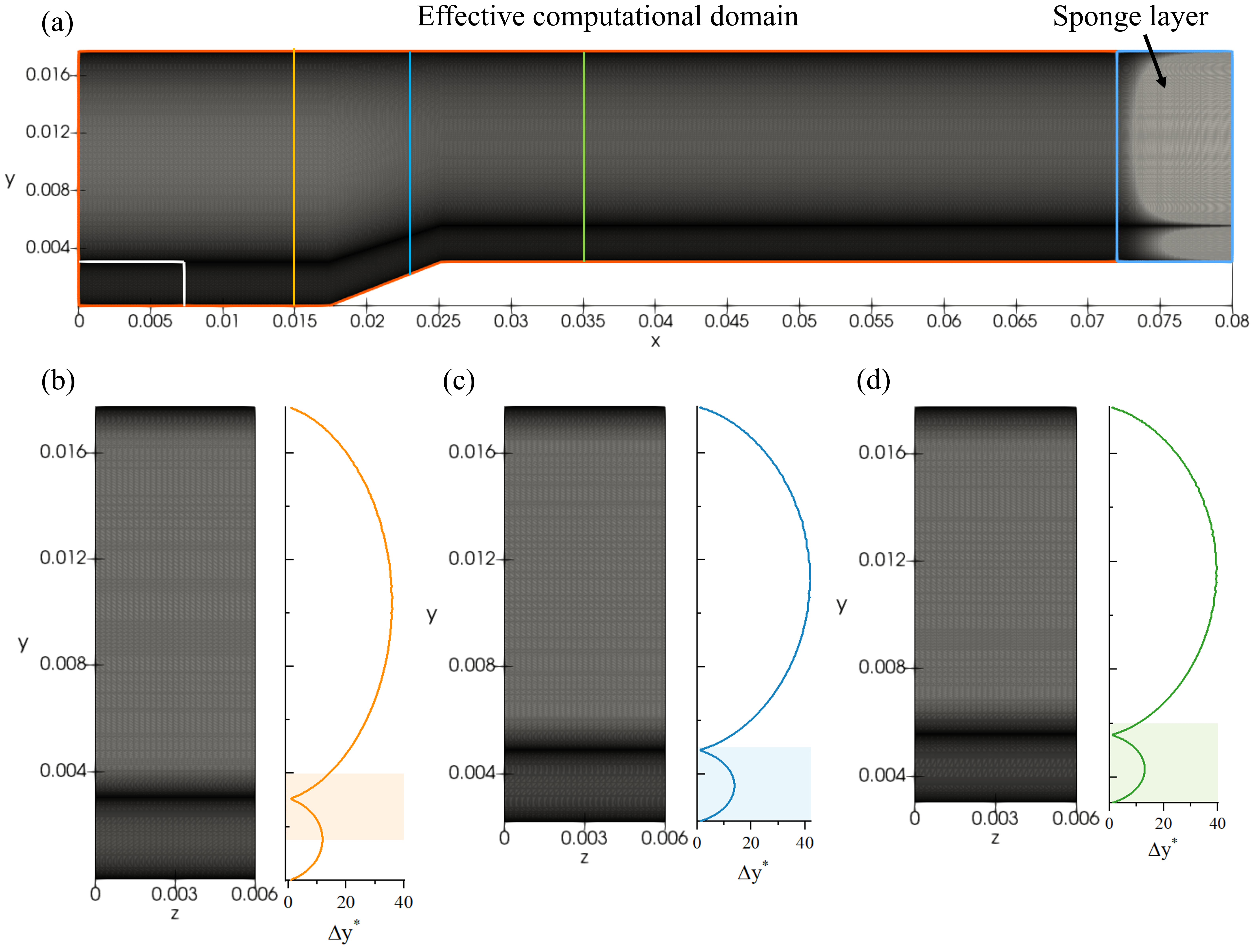}
    \fi
    \caption{(a): Mesh in $x-y$ plane with upstream wall and front wall of cavity delineated with white line; (b),(c) and (d): Mesh in $y-z$ plane at different streamwise location. Yellow, blue and green lines represent $x=15$~mm, $23$~mm and $35$~mm, separately. Light-colored area represents potential combustion region.} 
    \label{fig2:mesh}
\end{figure} 

Referring to the numerical simulation of Sitaraman et al.~\cite{SITARAMAN2021111531} and previous studies about combustion simulation under low pressure~\cite{Henrique2020104344}, the incoming freestream is set at 1000~K, 50060~Pa and a Mach number of 1.5. An equivalence ratio of 0.3 is selected in reactive simulation. Under this condition, the combustion near the cavity can establish within a reasonable runtime and also the fully burning state attain statistically stationary, which is essential for a detailed analysis. Corresponding to the above settings, the nominal thickness of turbulent boundary layer is $\delta_0$=1.47~mm and Reynolds number for the inlet flow is $Re_{\delta_0}$=5670.

\subsection{Mesh validation}
A non-uniform mesh with 1450 $\times$ 320 $\times$ 200 is used to discretise the domain in the streamwise, wall-normal, and spanwise directions, respectively. Figure.~\ref{fig2:mesh}(a) depicts the overall view of mesh distribution in $x-y$ plane. It can be seen that the mesh is smoothly stretched near the corner of the aft wall, with a refined resolution in the near-wall region. Within the effective computational domain, $\Delta x$ is of approximately 50~$\mu$m and $\Delta z$ has a constant value of 30~$\mu$m. Second row of Fig.~\ref{fig2:mesh} presents the variation of normalized size $\Delta y^*$ calculated by $\Delta y^* = \Delta y/min(\Delta y)$ at different locations. Profiles selected in Fig.~\ref{fig2:mesh} (b), (c) and (d) demonstrate mesh distribution in combustion region as flame propagates downstream. Minimum cells, which have length of about 4~$\mu$m in wall-normal direction, distribute at the wall and inside region around 3~mm from the bottom wall. 

\begin{table}[]
    \caption{Typical mesh size for DNS and LES of boundary layer flow.}
    \centering
    \label{table:1}
    \begin{tabular}{lll}
    \hline
    \textbf{}   & \textbf{DNS}& \textbf{wall-resolved LES}\\ \hline
     $ \Delta x^+ $          &   10-15   &   50-150            \\ 
    Min($ \Delta y^+ $)      &   1       &   1                 \\
    Points in $0 < y^+ < 10$ &   3       &   3                 \\
    $ \Delta z^+ $           &   5       &   10-20             \\ \hline
    \end{tabular}
\end{table}

To asses the quality of computational mesh, detailed validation are proposed in this section. Table~\ref{table:1} presents recommended resolution criterion of computational mesh cell sizes for both DNS and wall-resolved LES computations proposed by Sagaut~\cite{sagaut2007theoretical}. The present mesh characteristics calculated based on inlet profile are listed in Tables \ref{table:2}. The subscript '1' represents the value at the wall, considering the surface of the immersed boundary, while the subscript e' represents the value at the edge of the boundary layer. The effective mesh size, $\Delta e$, is defined as $\Delta e = \sqrt[3]{\Delta x \Delta y \Delta z}$. Results recommend that the DNS criteria are quite satisfactorily fulfilled in the present simulation, especially inside the reaction region. More details about mesh validation can be found in previous study~\cite{fang2023direct}.
\begin{table}[]
    \caption{Mesh validation at the inlet plane.}
    \centering
    \label{table:2}
    \begin{tabular}{lllllllllll}
    \hline
    $ \Delta x/{\eta}_1 $ & $\Delta y_1/{\eta}_1$ & $\Delta z/{\eta}_1$ & $\Delta e/{\eta}_1$  & $ \Delta x/{\eta}_e$ & $\Delta y_e/{\eta}_e$ & $\Delta z/{\eta}_e$ & $\Delta e/{\eta}_e$ & $ \Delta x^+ $ & $ \Delta y^+ $ & $ \Delta z^+ $\\ \hline
     7.69   & 0.60   & 4.51   & 2.75   & 1.78    & 2.61  & 1.04  & 1.69   & 12.10  & 0.95  & 7.10  \\ \hline
    \end{tabular}
\end{table}
In addition, Peterson~\cite{peterson2019high} proposed that simulations of a cavity flame-holder require a mesh resolution that satisfies a minimum of ten times the Kolmogorov scale in order to capture the key reacting flow structure. Results listed in Table~\ref{table:2} indicates that the present mesh can fully meets this requirement.
\begin{figure}[t]
    \centering\centering\ifx\mycmd\undefined
    \includegraphics[width=0.9\textwidth]{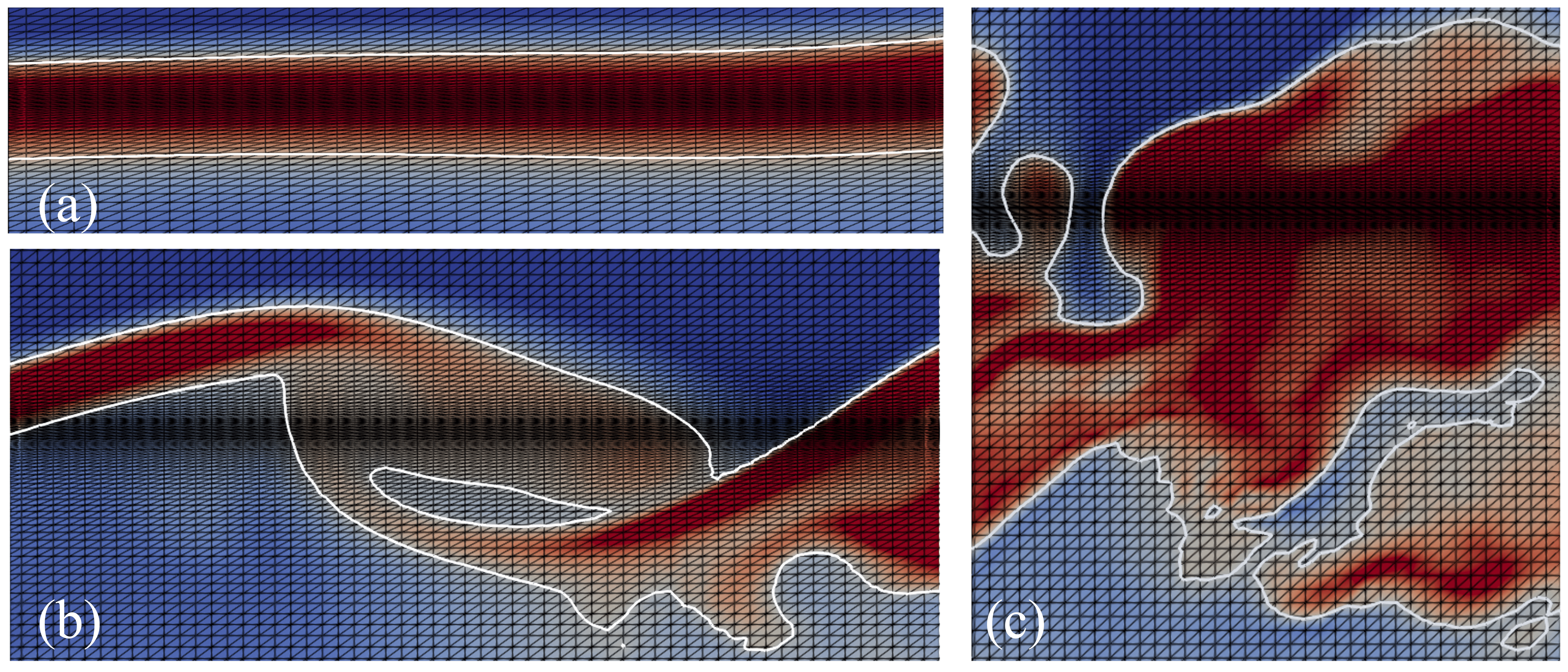}
    \fi
    \caption{Mesh in $x-y$ plane with counter plot of heat release rate, [J/m$^3$/s]. (a) near $x=$9~mm in Case 2; (b) near $x=$15~mm in Case 2; (c) near $x=$15~mm in Case 1.} 
    \label{fig3:mesh-Qdot}
\end{figure} 
\begin{table}[]
    \caption{Mesh validation via premixed counter flame thickness.}
    \centering
    \label{table:3}
    \begin{tabular}{llll}
    \hline
    flow condition  & flame thickness range & $n_{finest}$ & $n_{coarsest}$ \\ \hline
    0.5~atm, 1000~K         & 700~$\mu$m-1300~$\mu$m     & 30-72    & 16-30     \\ 
    0.5-0.8~atm, 1200~K     & 400~$\mu$m-700~$\mu$m      & 22-30    & 9-16      \\ \hline
    \end{tabular}
\end{table}
Moreover, based on the length scales calculated by Sitaraman et al.~\cite{SITARAMAN2021111531}, it can be observed that more than thirty of the finest cells, which are distributed near the boundary layer and inside the cavity, are included within the relevant length scales. To give a more detailed validation, flame thicknesses defined by $\delta_T = (T_{burnt}-T_0)/max(\nabla T)$ are calculated under different incoming velocity using one-dimensional premixed counter flame program and Cantera~\cite{cantera}. From range of incoming velocity under which reasonable result can be obtained from one-dimensional simulation, a series of flame thickness can be obtained. Therefore, Table~\ref{table:3} present the number of cells within flames under different condition. The effective cell size $\Delta e$ is used here to calculate $n_{finest}$ and $n_{coarsest}$. Coarsest cells in this table represent the coarsest cell size in reaction area marked in Fig.~\ref{fig2:mesh} (b), (c) and (d).
Figure.~\ref{fig3:mesh-Qdot} further depicts the cell distribution in reaction region. Above results and analysis all demonstrate that the present mesh is sufficiently refined to resolve the flow and capture the turbulent flame structure.

\section{Result and Discussion}
\label{sec:result}
For the inflow conditions considered in this study, the temperature near the separation corner can satisfied the condition for auto-ignition of the hydrogen-air mixture at the designated equivalence ratios. Thus, no extra treatment is required to ignite the flame. However, in order to avoid the high computational cost for evolving the reactive solutions from a \textit{zero} state, a sequence of steps are designed for the start-up of the reactive simulation.

The simulation was initiated with a non-reacting solution obtained from a 2D simulation with a laminar inflow. Two precursor DNS cases of channel flow were conducted to obtain the inflow data for three cases, respectively. An auxiliary DNS of a supersonic turbulent boundary layer flow was conducted to generate the inflow boundary-layer turbulence data, where the boundary layer transition was triggered using wall blowing and suction, and a series of flow slices were extracted and saved from the fully developed turbulent zone as the inflow data. The turbulent data was decomposed into mean flow and fluctuations. Uncorrelated fluctuations from two instants were mapped to the upper and lower boundary layers to form a fluctuation slice for Case 1. The temporal inflow fluctuations were then updated with the simulation using cubic spline interpolation of four continuous slices. More details and validations can be found in~\cite{fang2023direct}. As for Case 2, the inflow profiles were obtained via the laminar boundary layer analysis. The comparison of the inflow velocity profiles of these two cases is depicted in Fig.~\ref{fig1:geom}. It is shown that the skin friction at the inlet plane of Case 2 exhibits the same value as Case 1 but a notably reduced boundary layer thickness is observed for Case 2.

\begin{figure}[t]
    \centering\centering\ifx\mycmd\undefined
    \includegraphics[width=0.9\textwidth]{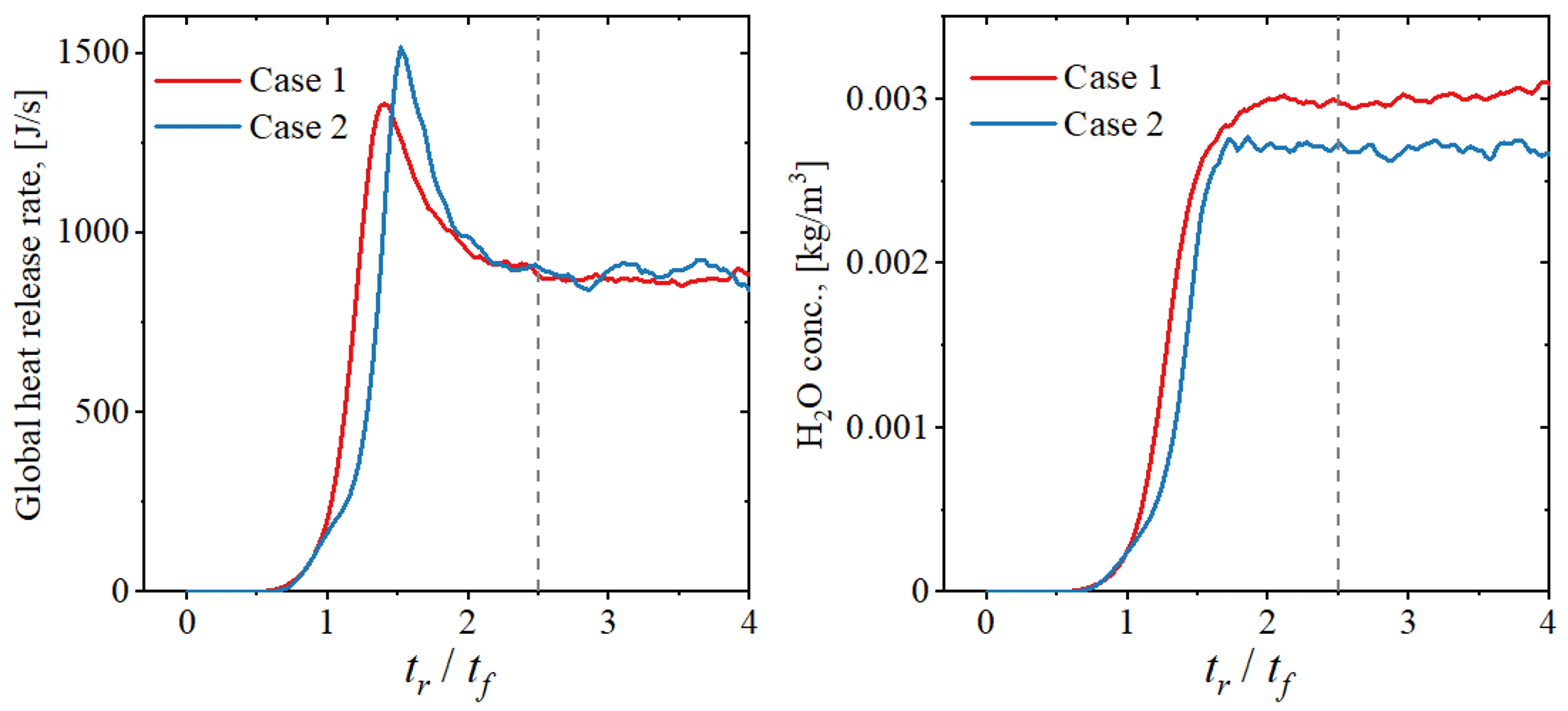}
    \fi
    \caption{Global heat release rate (left) and mean H$_2$O concentration (right) in near-cavity area. One flow-though-time $t_f = 7.57\times10^{-5}$ s, $t_r$ is the physical time counted since the reactant added in the domain.} 
    \label{fig4:steady}
\end{figure} 

All the simulations were monitored to ensure the flows reached statistically steady states, and afterwards data were collected to calculate statistics. With the addition of reactant in the domain, for all two cases combustion takes place spontaneously in the rectangular region near the cavity (4.5~cm in $x$-direction and 0.8~cm in $y$-direction, called \textit{near-cavity area} henceforth, marked in Fig.~\ref{fig1:geom}), and is able to reach a quasi-stationary state as indicated by the properties monitored in Fig.~\ref{fig4:steady}. A typical reactive flow simulation of five flow-through times required a wall-clock time of about 36 hours on 16384 cores.

\subsection{General steady-state features}
\begin{figure}[t]
    \centering\centering\ifx\mycmd\undefined
    \includegraphics[width=0.9\textwidth]{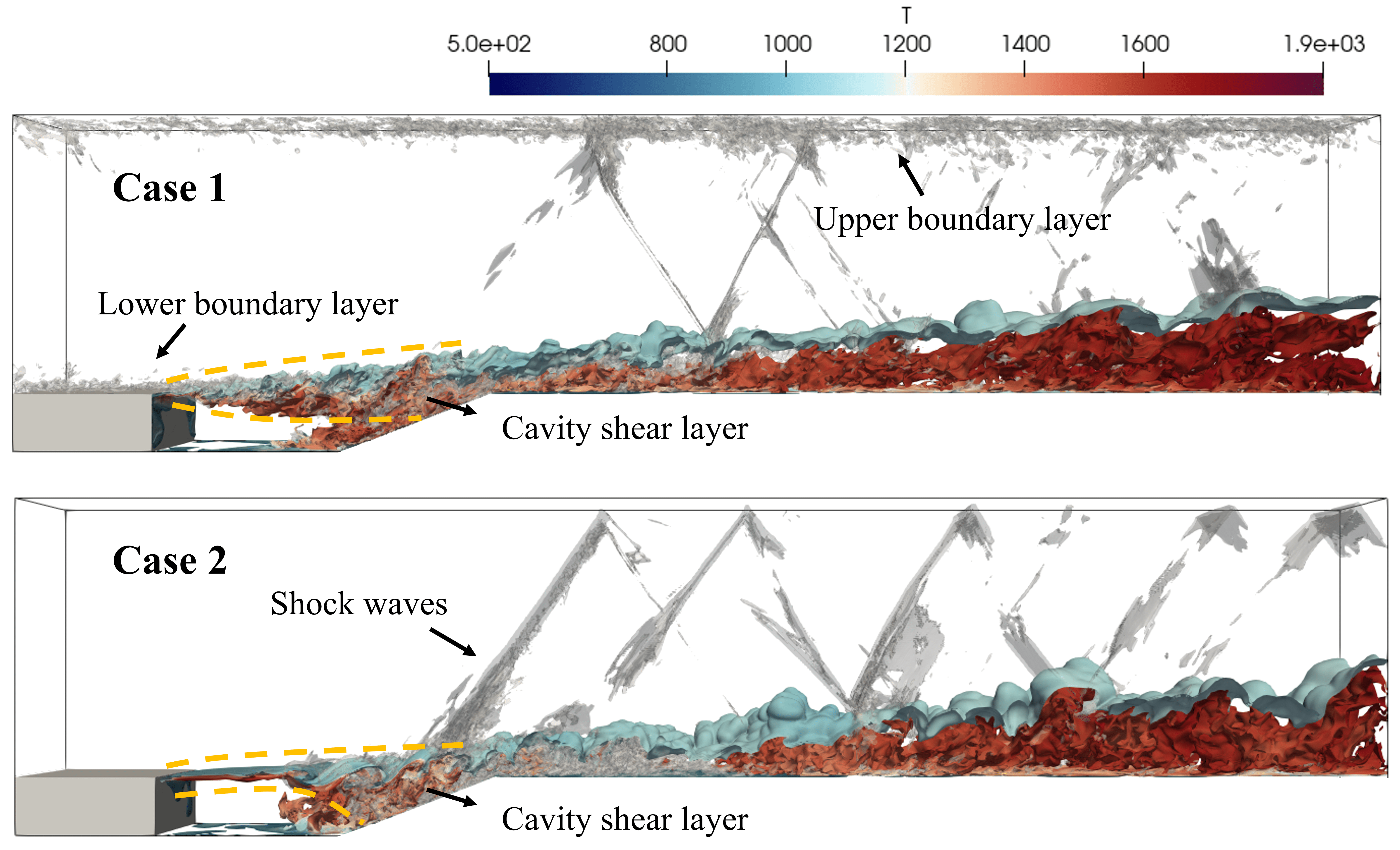}
    \fi
    \caption{Typical iso-surfaces of heat release rate and pressure gradient $|\nabla p|$ for two cases. Heat release rate is coloured by temperature. $|\nabla p|$ is contoured at the value of $2\times10^{7}$ Pa/m.} 
    \label{fig5:Qdot-P}
\end{figure}

To provide an overview of the flow field and flame morphology after the combustion in near-cavity area has reached the steady state, Fig.~\ref{fig5:Qdot-P} shows a three-dimensional representation of the near-cavity region using iso-surfaces of heat release rate (coloured by temperature) and pressure gradient magnitude (uncoloured) for two caes. Combustion takes place along the cavity shear layer and extends downstream along the lower wall. Within the cavity, the majority of the reaction occurs near the rear ramp, characterised by elevated heat release rate. According to the categorisation of combustion modes proposed by Wang et al.~\cite{WANG201312078}, the current condition can be classified as cavity shear-layer stabilised combustion. This mode of combustion is formed due to the elongated cavity and low equivalence ratio\cite{WANG201312078,BARNES201524}. 

\begin{figure}[t]
    \centering\centering\ifx\mycmd\undefined
    \includegraphics[width=0.9\textwidth]{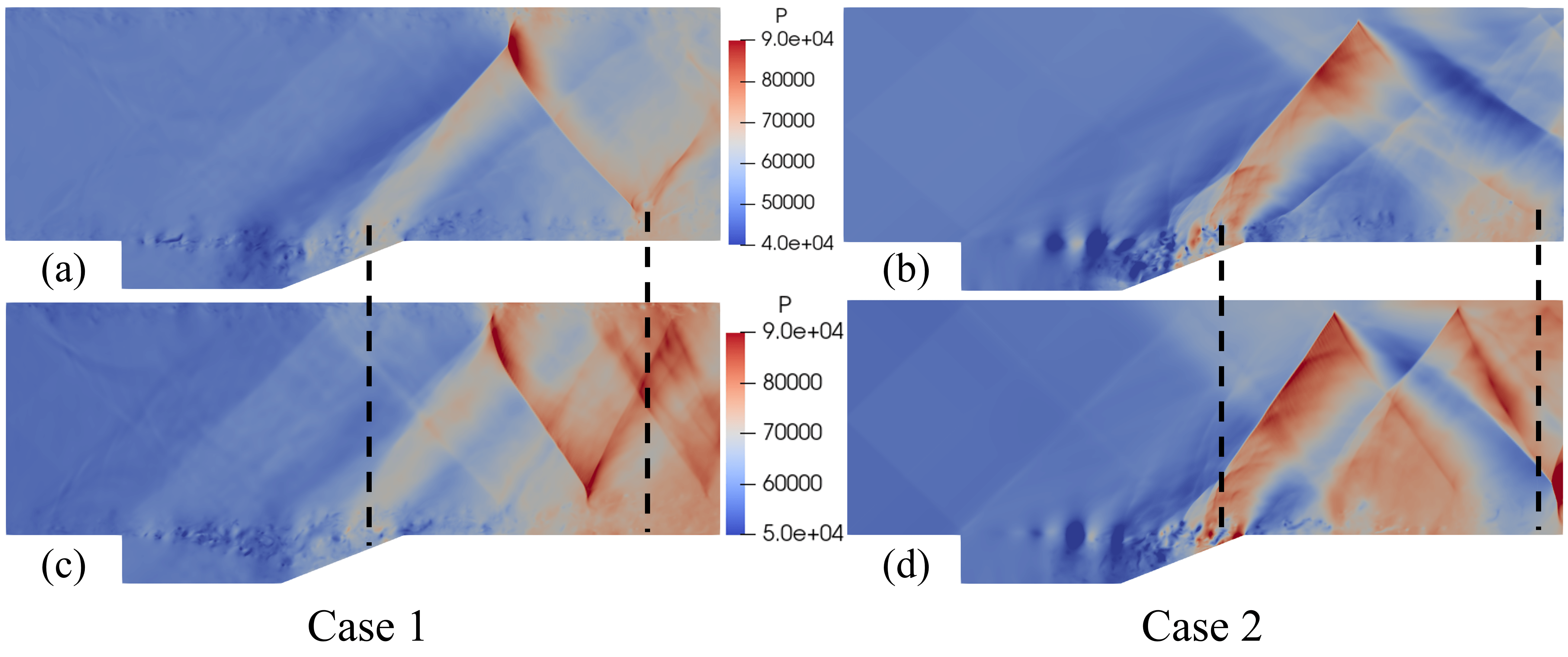}
    \fi
    \caption{The instantaneous plot of pressure in $x-y$ plane. To focus on the flow field of near-cavity area, the area with length of 4.5 cm and height of domain is presented. (a) non-reacting situation of Case 1; (b) non-reacting situation for Case 2; (c) reacting situation of Case 1; (d) reacting situation of Case 2.} 
    \label{fig6:P}
\end{figure} 

Additionally, the occurrence of combustion within the shear layer and boundary layer also leads to noticeable alterations in the flow field. The comparison of pressure distributions between the non-reacting and reacting conditions, depicted in Fig.~\ref{fig6:P}, is shown to examine the combustion effects. 

Results show that the impingement between the cavity shear layer and the aft wall leads to the formation of a compression wave. This compression wave moves towards upper wall and its intensity increase after the first SWBLI (shock-wave/boundary layer interaction) in Case 1 as depict in Fig.~\ref{fig6:P} (a), which is similar to the observation in the non-reactive condition~\cite{fang2023direct}. Meanwhile, Fig.~\ref{fig6:P} (b) shows that compression wave formed from the impingement point is observed to be stronger in Case 2 because of the more intense oscillation and become relatively weaker after reflected by the upper wall. Furthermore, due to the relatively high temperature inside the cavity with reaction, weak compression waves observed in Fig.~\ref{fig6:P} (c) and (d) originate from the separation corner and undergo successive reflections between the upper and lower walls. After reflected by the upper wall, it interacts with the lower boundary layer and flame then become visibly stronger in both cases. The reason for this may be that the interaction between the compression wave and reactive turbulent boundary layer gives rise to a strengthened wave. Similar behaviour can be observed also for the stronger shock front interacting with the lower boundary layer.

In addition, considering the impingement point on the rear wall and the acting point of the reflected shock wave on the lower wall, it is observed that the distance between these two points is shorter in the reactive case. This phenomenon can be attributed to the combustion occurring in the lower boundary layer, which cause elevated back pressure and consequently lead to the a significant compression of the compression waves.

\begin{figure}[ht]
    \centering\centering\ifx\mycmd\undefined
    \includegraphics[width=0.6\textwidth]{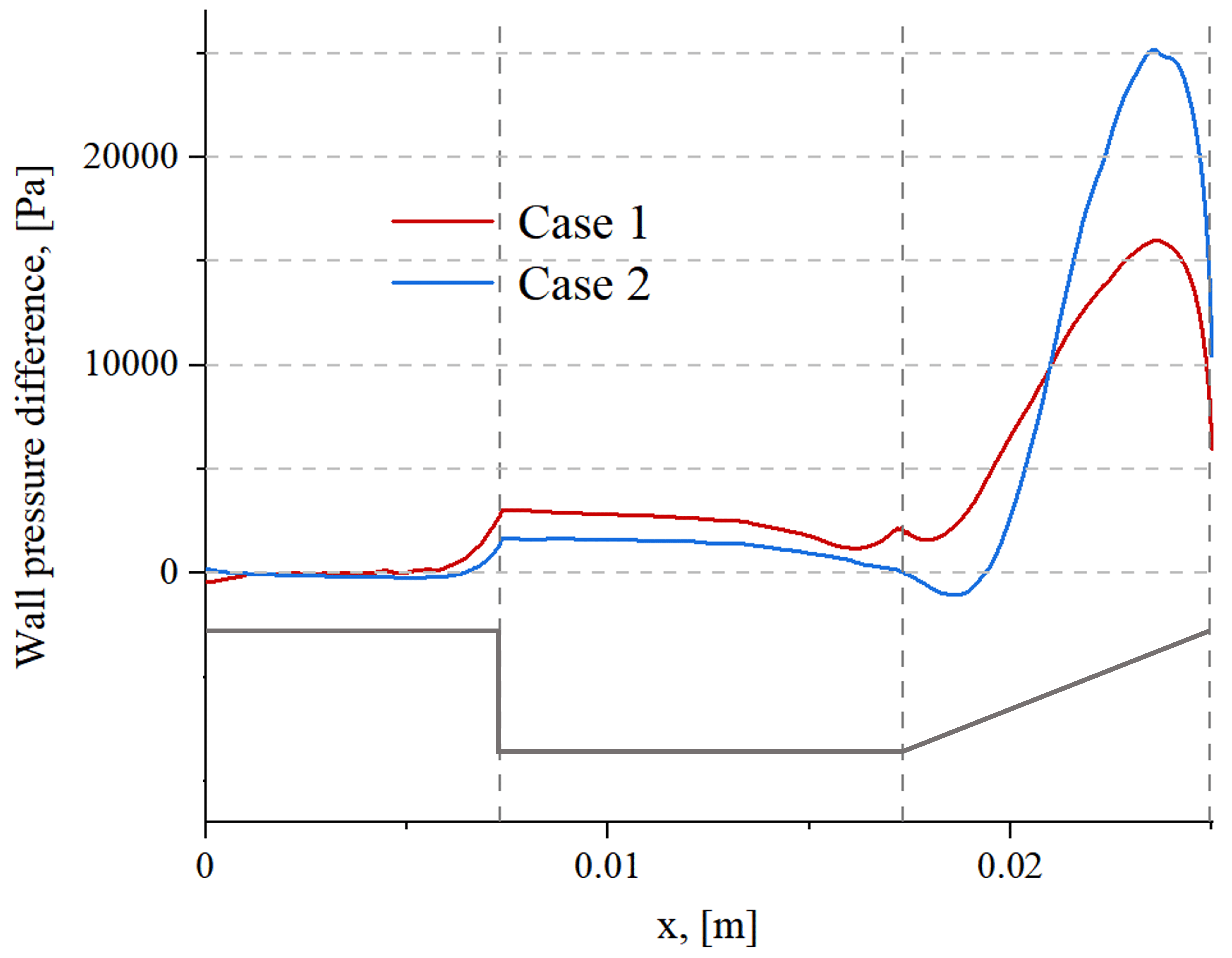}
    \fi
    \caption{Difference of time-averaged wall pressure that integrated in spanwise direction. Wall pressure difference is calculated by local wall static pressure minus averaged upstream wall static pressure.} 
    \label{fig7:wallP}
\end{figure}

When designing scramjets, aerodynamic drag plays a crucial role that necessitates careful consideration~\cite{gruber2004mixing}. Previous studies have indicated that pressure drag constitutes the dominant component of cavity drag~\cite{vortex}. To assess the drag in the three cases, a comparison of wall pressures is conducted. Figure~\ref{fig7:wallP} illustrates the axial variations in the wall pressure, where the influence of slightly different upstream pressures is neglected. In comparison to the outcomes observed in Case 1, pressure difference rises more significantly along the aft wall and reaches a higher peak value in Case 2. This finding further confirms the occurrence of intensified collision and higher pressure drag in Case 2. 

\subsection{Mass transport process}
\begin{figure}[t]
    \centering\centering\ifx\mycmd\undefined
    \includegraphics[width=0.8\textwidth]{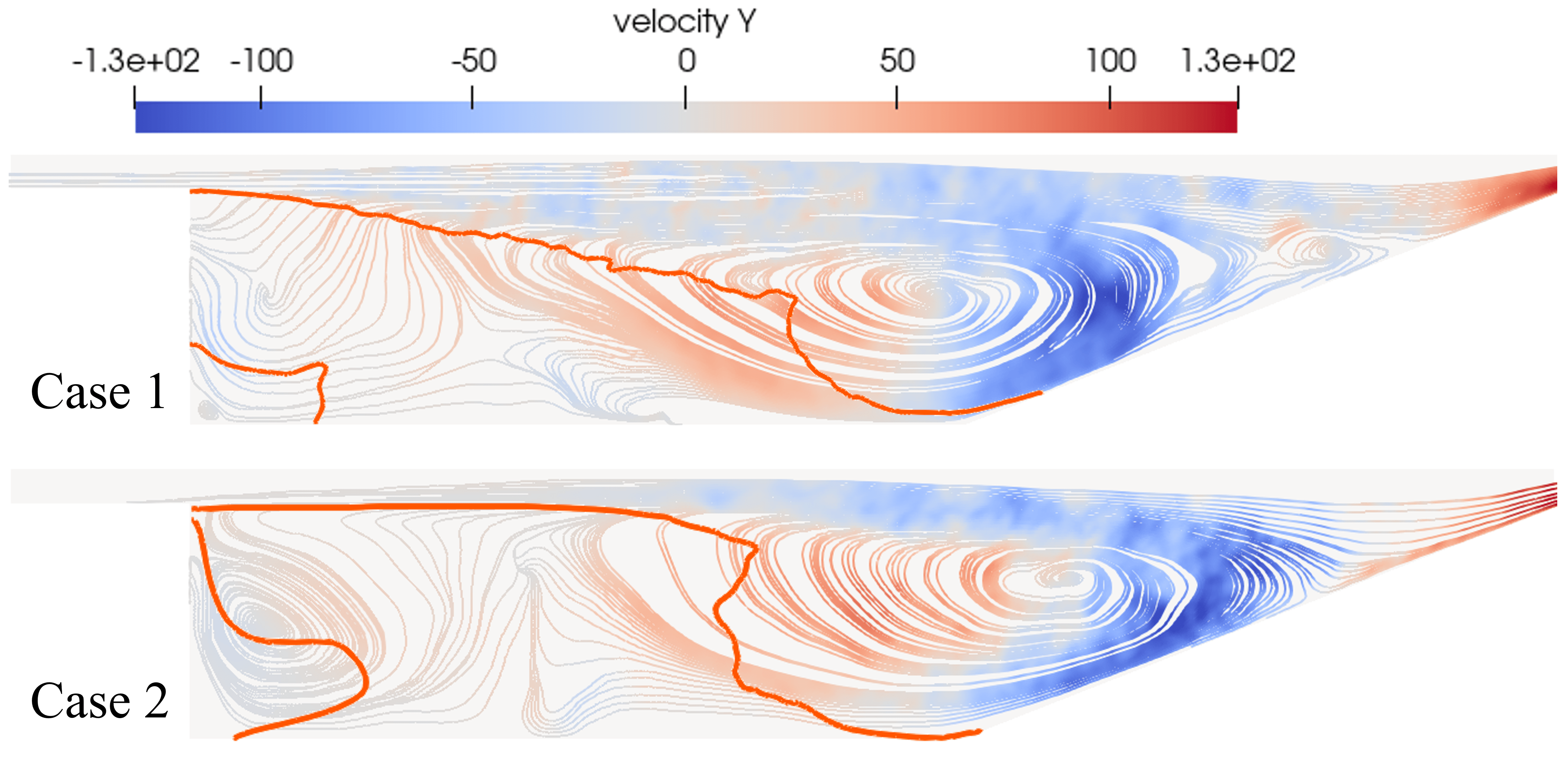}
    \fi
    \caption{Time-averaged streamlines in $x-y$ plane colored by velocity component along the y direction, [m/s]. Red line is the contour line of Y$_{H_2O}$ at the value of $0.071$.} 
    \label{fig8:streamwise}
\end{figure}

It is hard to validate the mass exchange level of  Case 1 and Case 2 only based on the observation of the instantaneous flow field. Therefore Fig.~\ref{fig8:streamwise} depicts the main recirculation zone in the cavity via time-averaged streamlines. Two main vortices are observed inside the cavity and primary vortex mainly distributes near the aft wall. Results depicted in Fig.~\ref{fig8:streamwise} indicate that the primary vortex size is marginally reduced in Case 2, whereas the primary vortex depth is slightly greater in Case 1. However, the distinctions between the primary vortices of Case 1 and Case 2 are not readily discernible.

Regarding the anterior region of the cavity, the primary mechanism responsible for the transport of burnt gas and radicals is the secondary vortex. Validation between the secondary vortices of Case 1 and Case 2 exhibits that their strengths are highly comparable. However, the interaction between primary and secondary vortex is notably weaker in Case 2 and a barrier-like structure is observed in the middle of the cavity. Therefore, additional evidence is required to facilitate the analysis.

\begin{figure}[t]
    \centering\centering\ifx\mycmd\undefined
    \includegraphics[width=0.9\textwidth]{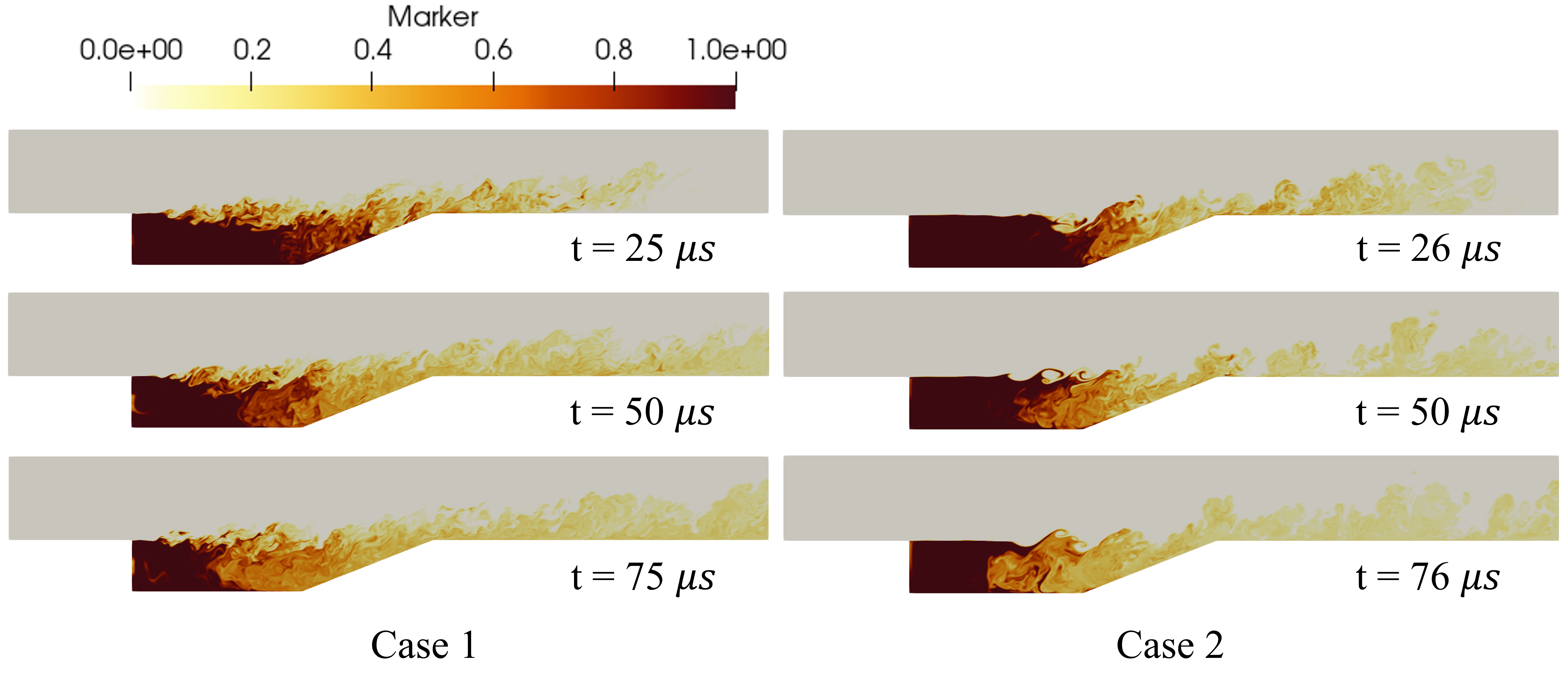}
    \fi
    \caption{Snapshots of the time evolution of cavity mass transport in $x-y$ plane. Consider t = 0 $\mu$s is the moment that gas in the cavity is marked.} 
    \label{fig9:marker}
\end{figure} 

In an attempt to obtain detailed analysis of mass exchange between the cavity and the main stream, the following method is adopted following the previous studies of Baurle et al.~\cite{baurle2000analysis} and Gruber et al.~\cite{vortex}. After the reactant was added in the domain, according to the results in Fig.~\ref{fig4:steady}, the simulation was advanced for about 2.5 flow-through-time to let the combustion in the near-cavity area reach statistically steady state. Then a marker scalar is added and transported in the simulation. The primary value of the marker is set to be one and zero for gas within and outside the cavity, respectively. It could be considered as a specially designed component that would only be transported by convection since the molecular diffusion is neglectable in this process, therefore diffusion term is disabled for the marker transport equation. This marker, in some cases, can also be regarded as \textit{residence time} tracer. 

Figure~\ref{fig9:marker} shows the time evolution of the mass exchange and transport processes using the marker field snapshots. It is shown that for Case 1, the main stream fluid entering the cavity distributes more uniformly over a wider region. Conversely, for Case 2, the entered gas primarily accumulates in the back part of the cavity shear layer and get transported into the central region of cavity via primary vortex. This observation, in conjunction with the findings presented in Fig.~\ref{fig8:streamwise}, reveals that the region where mass exchange predominantly occurs remains consistent with the location of the primary vortex. Furthermore, results reveal high level of mass exchange in the leading portion of cavity shear layer in Case 1. In contrast, mass transfer process between inner and outer gas appears to be negligible in the anterior region of cavity in Case 2. Observation confirms that the secondary vortex in Case 2 primarily facilitates the internal gas transport within the cavity because of the undisturbed cavity shear layer on top of this region. Also, this undisturbed part of cavity shear layer partially account for the weaker combustion observed in the leading edge of the shear layer in Fig.~\ref{fig5:Qdot-P}.

Moreover, recirculation process induced by primary vortex and secondary vortex appears to be more independent from each other in Case 2 because of the barrier-like structure observed in Fig.~\ref{fig8:streamwise}, leading to the limited ability of secondary vortex to transport products out of the anterior section of the cavity. This phenomenon could contribute to the deposition of reaction products plotted in Fig.~\ref{fig8:streamwise}. Additionally, the lower mass flux and velocity in this region suggest an extended residence time, facilitating more complete reactions.

To quantify the mass exchange for the cavity, the mass flow across the interface between the cavity and main flow is calculated using:
\begin{equation}
\phi_{in} = \sum\limits_{i,v_i<0} \rho_{i} \cdot v_{i} \cdot s_{i}\:,
\end{equation}
\begin{equation}
\phi_{out} = \sum\limits_{i,v_i>0} \rho_{i} \cdot v_{i} \cdot s_{i}\:,
\end{equation}
where $\phi_{in}$ and $\phi_{out}$ stand for mass flow into and out of cavity, respectively. $\rho_{i}$ and $v_{i}$ represent the density and the velocity component along the $y$ direction of the $i$-th point on the interface. $s_{i}$ is the size of area formed by point $i$ and its adjacent points. 

\begin{table}[]
    \caption{Data of time-averaged mass flow into and out of cavity.}
    \centering
    \label{table4:exchange}
    \begin{tabular}{lll}
    \hline
    \textbf{Case} &  \textbf{Mass flow into cavity, [kg/s]}    & \textbf{Mass flow out of cavity, [kg/s]} \\ \hline
    Case 1     &     $1.427\times10^{-4}$            &    $1.294\times10^{-4}$         \\         
    Case 2     &     $1.748\times10^{-4}$            &    $1.592\times10^{-4}$         \\ \hline
    \end{tabular}
\end{table}

Table~\ref{table4:exchange} lists the result obtained using mean density and velocity in both cases. It is interesting to see that mass flow rates for both inflow and outflow are considerably higher in Case 2. Combining with the observation made from Fig.~\ref{fig9:marker}, it can be speculated that the primary vortex near the aft wall bring a significant enhancement of the mass exchange between the cavity and main stream in Case 2. The occurrence of this phenomenon can be attributed to the advanced collision at the reattachment point.

In summary,  Case 2 demonstrates a stronger mass exchange process within the rear cavity region, whereas the enhanced shear layer generated by inflow wall-bounded turbulence promotes a more uniformly distributed mass exchange process in Case 1. Because of the barrier-like structure observed in Case 2, the transport of gas in the anterior cavity region is impeded, indicating an extended residence time and stronger production deposition. Moreover, the undisturbed shear layer above the anterior part of cavity in Case 2 is unsuitable for facilitating reactions.

\subsection{flame stretch and thickness}
To shed more light into the turbulence-flame interaction, it is of an interest to examine the flame stretch and thickness under the supersonic combustion conditions. Progress variable is calculated based on the mass fraction of H$_2$O using
\begin{equation}
c = \frac{Y_{H_2O}}{Y^b_{H_2O}} \:,
\end{equation}
where the subscript `b' represents the burnt mass fraction at an initial temperature of 1000 K and pressure of 0.5 atm.
\begin{figure}[t]
    \centering\centering\ifx\mycmd\undefined
    \includegraphics[width=0.8\textwidth]{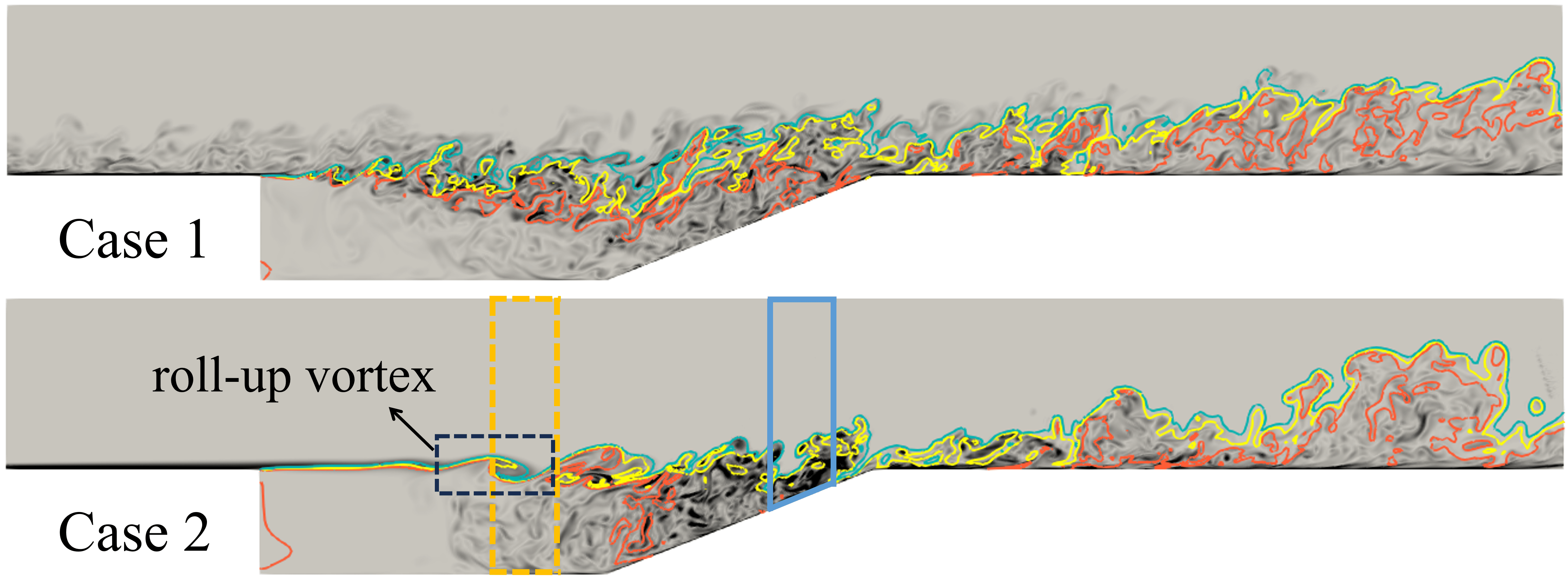}
    \fi
    \caption{Instantaneous plot of contour line of progress variable (green: 0.2, yellow: 0.5, orange: 0.8) with contour plot of vorticity in background in $x-y$ plane of near-cavity area. Region near $x=15$~mm and $x=23$~mm is marked by yellow dot box and blue solid box separately.} 
    \label{fig10:c_vor}
\end{figure} 
\begin{figure}[t]
    \centering\centering\ifx\mycmd\undefined
    \includegraphics[width=0.7\textwidth]{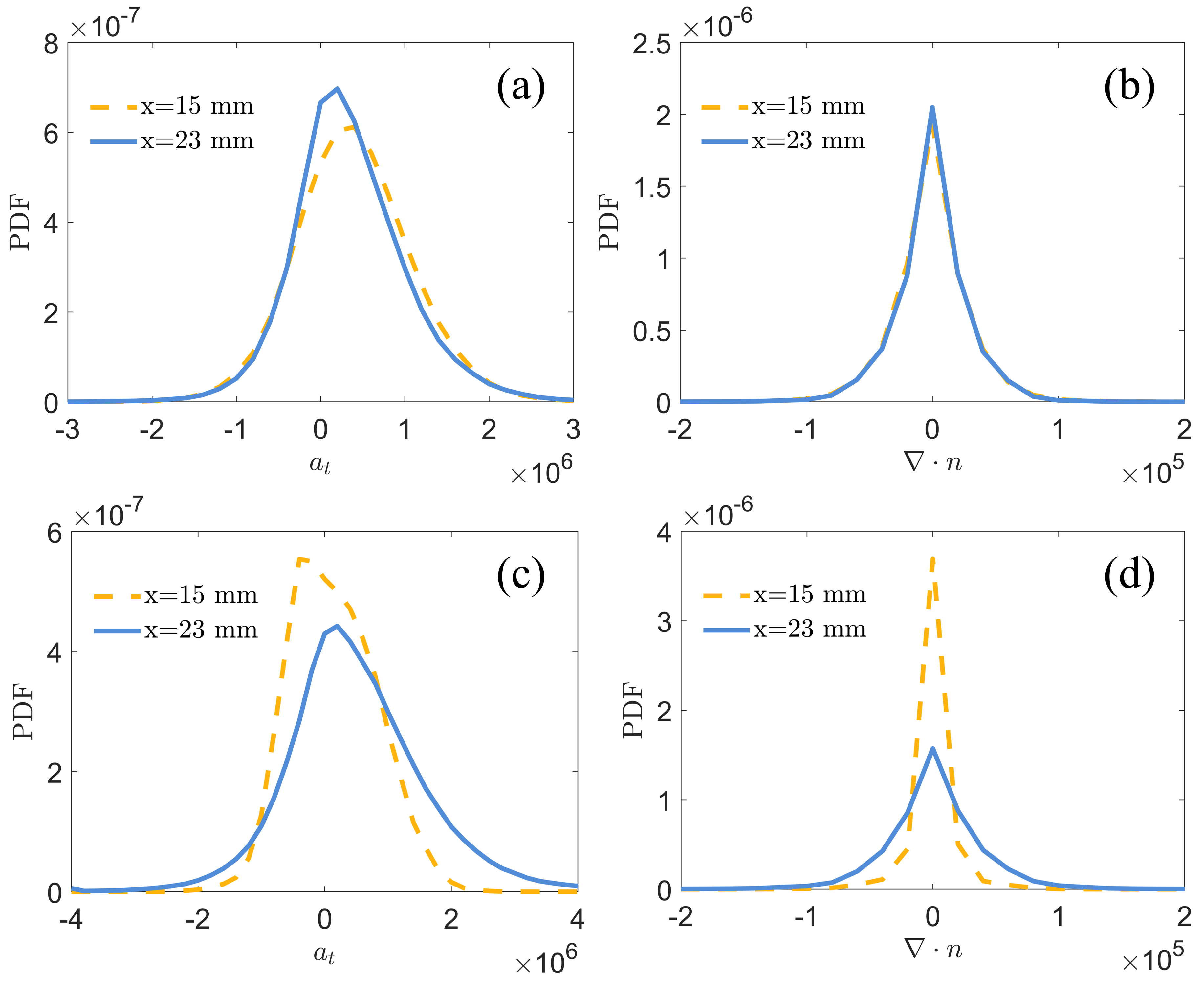}
    \fi
    \caption{PDFs of tangential strain rate (left) and the curvature (right) in reaction zone (0.1$<$c$<$0.8) at different $x$ locations. (a) and (b): Case 1; (c) and (d): Case 2.} 
    \label{fig11:atcur_pdf}
\end{figure} 
Figure~\ref{fig10:c_vor} depicts the distribution of progress variable iso-contours overlaid with the vorticity field within the cavity shear layer and boundary layer.  It can be seen that vortices predominantly lies within the expansive oxidation zone (indicated by the three progress variable iso-values), particularly in the vicinity near the collision point. In both Case 1 and Case 2, the oxidation layer grows wider as the vortices gradually develop along the cavity shear layer. Especially in Case 2, the thickness of oxidation layer begins to increase significantly owing to the impingement of the reactive shear layer onto the rear wall. Furthermore, a pronounced roll-up vortex is detected in region close to $x=15$~mm in Case 2. Its influence on flame structure is validated in the following discussion.

To further analysis this interaction,flame stretch, curvature $\kappa$ and tangential strain rate $a_t$ are calculated as follows:
\begin{equation}
\kappa = \nabla \cdot \textbf{n},
\end{equation}
\begin{equation}
    a_t = \nabla \cdot \textbf{u} - n_i S_{ij} n_j,
\end{equation}
\begin{equation}
\frac{\dot{\delta A}}{\delta A} = a_t + S_d \nabla \cdot \textbf{n},
\end{equation}
where $\frac{\dot{\delta A}}{\delta A}$ represent the flame stretch, $\textbf{n} = -\nabla c/|\nabla c|$ is the flame surface normal vector, $S_{ij}=\frac{1}{2}(\frac{\partial u_i}{\partial x_j}+\frac{\partial u_j}{\partial x_i})$ is the rate-of-strain tensor and $S_d=\frac{\dot{\omega_c}}{\rho |\nabla c|} + \frac{1}{\rho |\nabla c|}\frac{\partial}{\partial x_i}(\rho D_c \frac{\partial c}{\partial x_i})$ is the progress variable isosurface propagation speed.

Figure~\ref{fig11:atcur_pdf} presents the PDF of the tangential strain rate and curvature at various downstream locations, representing different evolution stages of the cavity shear layer. According to the results depicted in Fig.~\ref{fig11:atcur_pdf} (a) and (c), relatively broad PDFs of $a_t$ peak at positive value in both cases. As the flame propagate downstream, the width of $a_t$ PDF slightly decrease in Case 1, which is consistent with the observation for previous study~\cite{Chakraborty2007}. As for the flame curvature PDFs, identical behaviours are observed in Fig.~\ref{fig11:atcur_pdf} (b) along the axial locations in Case 1. This phenomenon is similar to that observed in the DNS of a premixed jet flame~\cite{wang2017direct} having a similar flame evolution in the jet shear layer. In addition, this trend seems to be different in Case 2. Compared with PDFs in Case 1, the PDF of tangential strain rate exhibits broader distribution range near $x=23$~mm and curvature PDF demonstrates a comparatively reduced spread close to $x<15$~mm. 
\begin{figure}[t]
    \centering\centering\ifx\mycmd\undefined
    \includegraphics[width=0.7\textwidth]{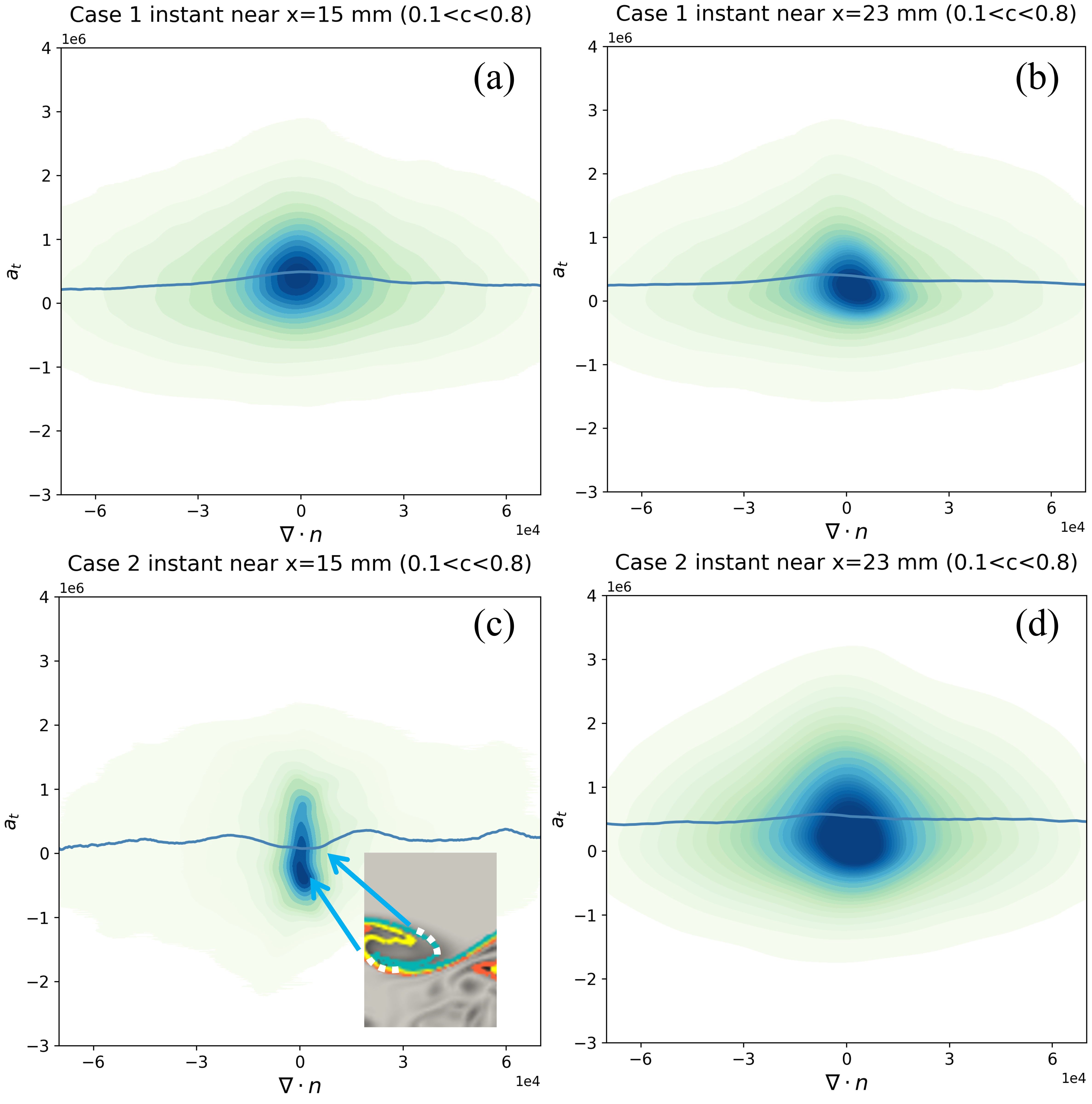}
    \fi
    \caption{Joint PDF of the tangential strain rate and the curvature in reaction region ($0.1 < c < 0.8$) at different x locations. The conditional mean is marked by the blue solid line.} 
    \label{fig12:cur_at}
\end{figure} 
The observed phenomenon could be attributed to the incompletely developed turbulence within the anterior region (i.e. $x<15$~mm) of the cavity and a sudden growth of vorticity at the impingement point (see Fig.~\ref{fig10:c_vor}). Also, pronounced roll-up vortex exists near $x<15$~mm. These are reflected here via the significant change in PDFs at the $x=15$~mm and $23$~mm locations for Case 2.

\begin{figure}[t]
    \centering\centering\ifx\mycmd\undefined
    \includegraphics[width=0.7\textwidth]{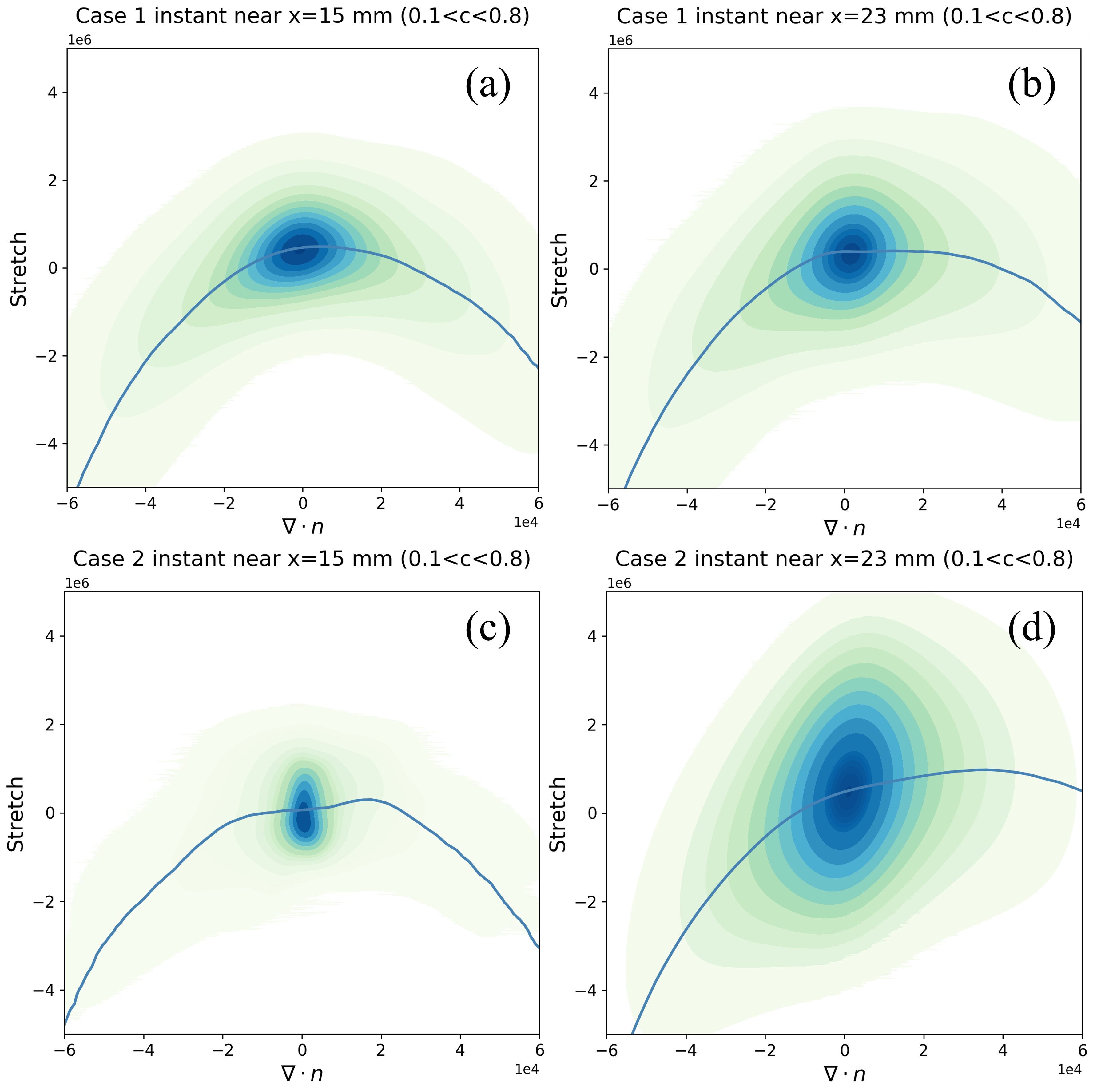}
    \fi
    \caption{Joint PDF of the flame stretch and the curvature in reaction region ($0.1 < c < 0.8$) at different x locations. The conditional mean is marked by the blue solid line.} 
    \label{fig13:cur_stretch}
\end{figure} 

Further analysis is conducted to elucidate the impact of tangential strain rate $a_t$ and curvature $\nabla \cdot n$. Joint PDF of tangential strain rate and curvature is initially investigated in Fig.~\ref{fig12:cur_at}. The findings suggest that positive averaged tangential strain rate $a_t$ is present across nearly all regions. Conditional averaged $a_t$ reach slightly elevated level in regions characterized by low curvature magnitudes in Case 1 near $x=15$~mm, which is consistent with the result of jet flame~\cite{wang2017direct}. 

Moreover, it is interesting to discover that tangential strain rate maintains a consistently low magnitude because of the transitioning turbulence. Also, a comparatively decreased $a_t$ occurs in the region that have low curvature magnitude near $x=15$~mm in Fig.~\ref{fig12:cur_at} (c). 
\begin{figure}[t]
    \centering\centering\ifx\mycmd\undefined
    \includegraphics[width=0.7\textwidth]{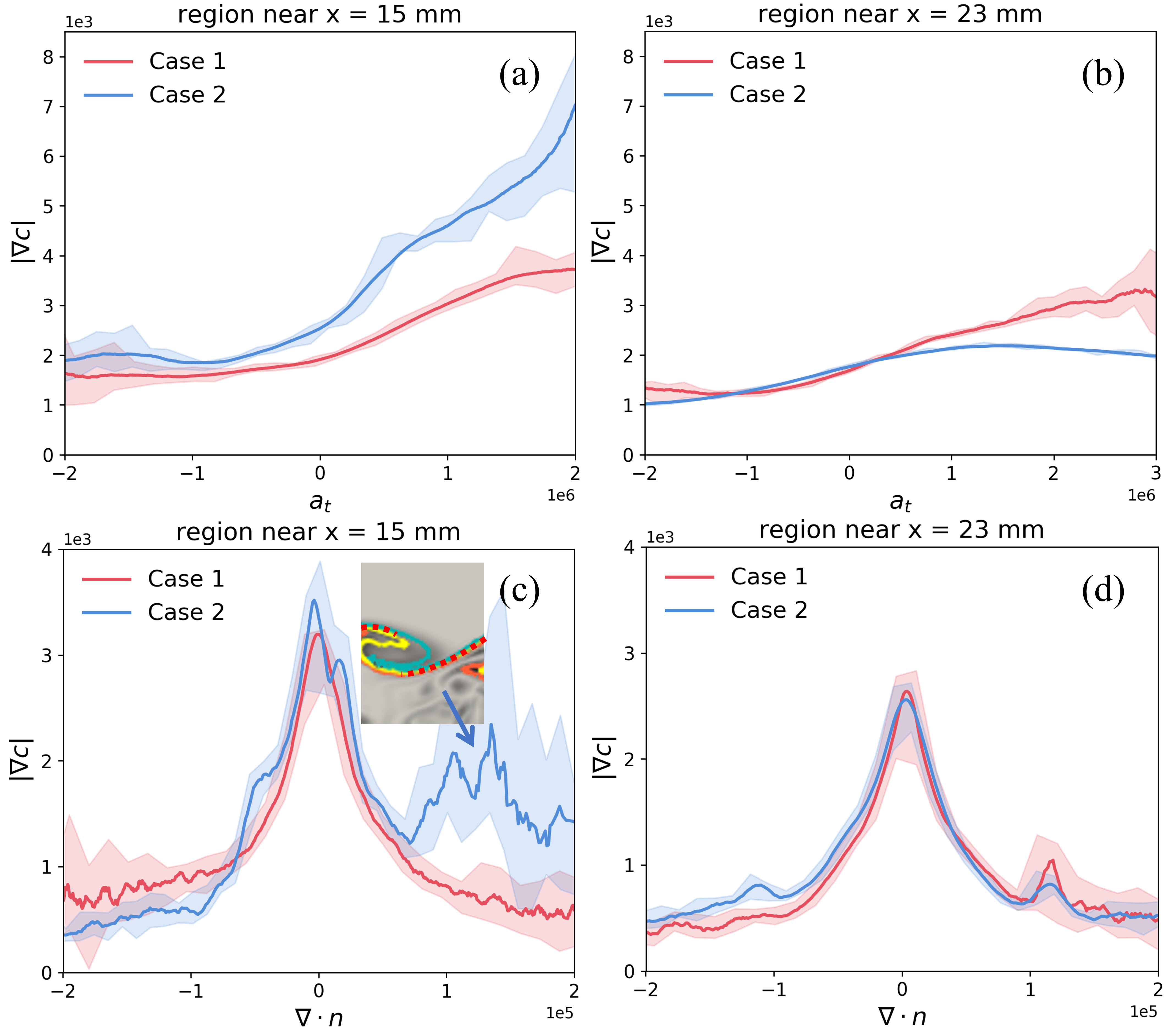}
    \fi
    \caption{(a) and (b): $|\nabla c|$ conditionally averaged on tangential strain rate $a_t$; (c) and (d): $|\nabla c|$ conditionally averaged on curvature $\nabla \cdot n$. Lines are smoothed results and colored areas are distribution range of conditional means before smoothing.} 
    \label{fig14:gradC_atcur}
\end{figure} 
This phenomenon can be ascribed to the presence of the roll-up vortex. The flame rolled by this vortex exhibits wrinkled surface and demonstrates negative tangential strain rate at the tip of roll-up vortex marked by white dot line in Fig.~\ref{fig12:cur_at} (c). Figure.~\ref{fig13:cur_stretch} further illustrates the joint PDF between flame stretch and curvature. Results indicate that stretch reach its maximum value in regions characterized by low curvature magnitudes and decrease as curvature magnitude increase in Case 1. Additionally, in Case 2 near $x=15$~mm, stretch is observed to be lower in proximity to $\nabla \cdot n = 0$. This observation could be attributed to the decreased tangential strain rate observed in Fig.~\ref{fig12:cur_at} (c) in this specific region. When comparing the outcomes depicted in Fig.~\ref{fig13:cur_stretch} (b) and (d) with (a) and (c), stretch presented in Fig.~\ref{fig13:cur_stretch} (b) and (d) observed to have slightly higher values in regions exhibiting high positive curvature near $x=23$~mm, attributed to the collision at the rear wall. The interaction with wall might could significantly influence the progress variable isosurface propagation speed $S_d$ and consequently lead to increased stretch. This phenomenon is particularly evident in Case 2 and positive flame stretch is observed.


The analysis of turbulence on flame thickness is further conducted. In region around $x=15$~mm, it is evident from Fig.~\ref{fig14:gradC_atcur} (a) that $|\nabla c|$ maintains a low value and displays reduced sensitivity to negative $a_t$; but as $a_t$ transitions to positive values, $|\nabla c|$ correspondingly rises and lead to reduction in flame thickness. This result is consistent with DNS study of Wang et al.~\cite{wang2017direct}. In contrast to Case 1, decreased flame thickness is observed in this region in Case 2, which show great agreement with results depicted in Fig.~\ref{fig10:c_vor}. Based on the findings illustrated in Fig.~\ref{fig14:gradC_atcur} (b) where the flame collides with the aft wall, the flame thickness increases in both cases and exhibits decreased sensitivity to changes of tangential strain rate. Furthermore, $|\nabla c|$ converges to a more comparable extent in both cases and a marginal smaller $|\nabla c|$ is noted within regions displaying large positive $a_t$ in Case 2. 

As for the influence of curvature, the results depicted in Fig.~\ref{fig14:gradC_atcur} (c) and (d) present a pronounced reduction in flame thickness with lower curvature magnitude, which is consistent with the direct numerical simulation by Sankaran et al.~\cite{SANKARAN20071291} and the experimental results by Tamadonfar and Gülder~\cite{tamadonfar2015experimental}. Furthermore, according to the results depicted in Fig.~\ref{fig14:gradC_atcur} (c), relatively large $|\nabla c|$ is evident in regions featuring pronounced positive curvature. This occurrence could be attributed to the influence of the roll-up vortex. Positive curvature is observed within the inner layer of roll-up vortex marked by red dot line. Simultaneously, elevated positive tangential strain rate is also noted within this zone, contributing to the stretch of this portion of flame. Near $x=23$~mm, with the the dissipation of roll-up vortex, $|\nabla c|$ reduces and sustains a diminished magnitude with the growth of positive curvature. 

Overall, consistent with the preceding results~\cite{wang2017direct}, the stretching impact of the tangential strain rate acts to reduce flame thickness, while curvature with large magnitude likely to dissipate the scalar gradient and thus thicken the flame. The inlet wall turbulence influence this process mainly by affecting turbulence evolution.
\begin{figure}[t]
    \centering\centering\ifx\mycmd\undefined
    \includegraphics[width=0.9\textwidth]{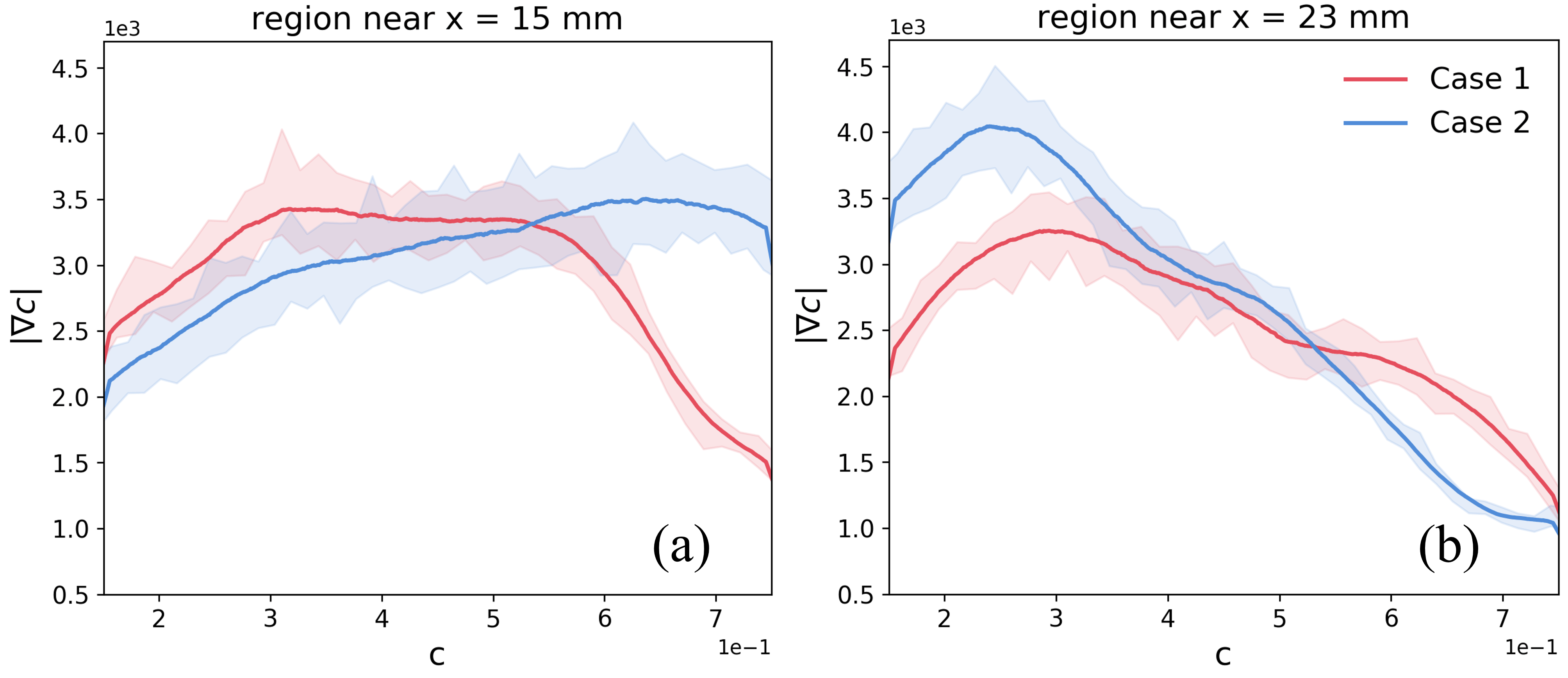}
    \fi
    \caption{$|\nabla c|$ conditionally averaged on progress variable c. Lines are smoothed results and colored areas are distribution range of conditional means before smoothing.} 
    \label{fig15:gradC_c}
\end{figure} 

To delve deeper into the underlying factors contributing to the variation of flame thickness in two cases, Fig.~\ref{fig15:gradC_c} presents conditional averages of $|\nabla c|$ at various locations over the process variable c. As the flame progresses over a distance ($x$ near $15$~mm), flames in two cases exhibit comparable thicknesses in the preheated area characterized by low c. With the increase of progress variable, the flame thickness in Case 1 undergoes significant growth in oxidation area and become noticeably larger than that in Case 2. After that, upon the flame impinging on the step, flames in both cases demonstrate a marked increase in thickness in wider range of progress variable. These findings are consistent with the results depicted in Fig.~\ref{fig10:c_vor}.
\begin{figure}[t]
    \centering\centering\ifx\mycmd\undefined
    \includegraphics[width=0.9\textwidth]{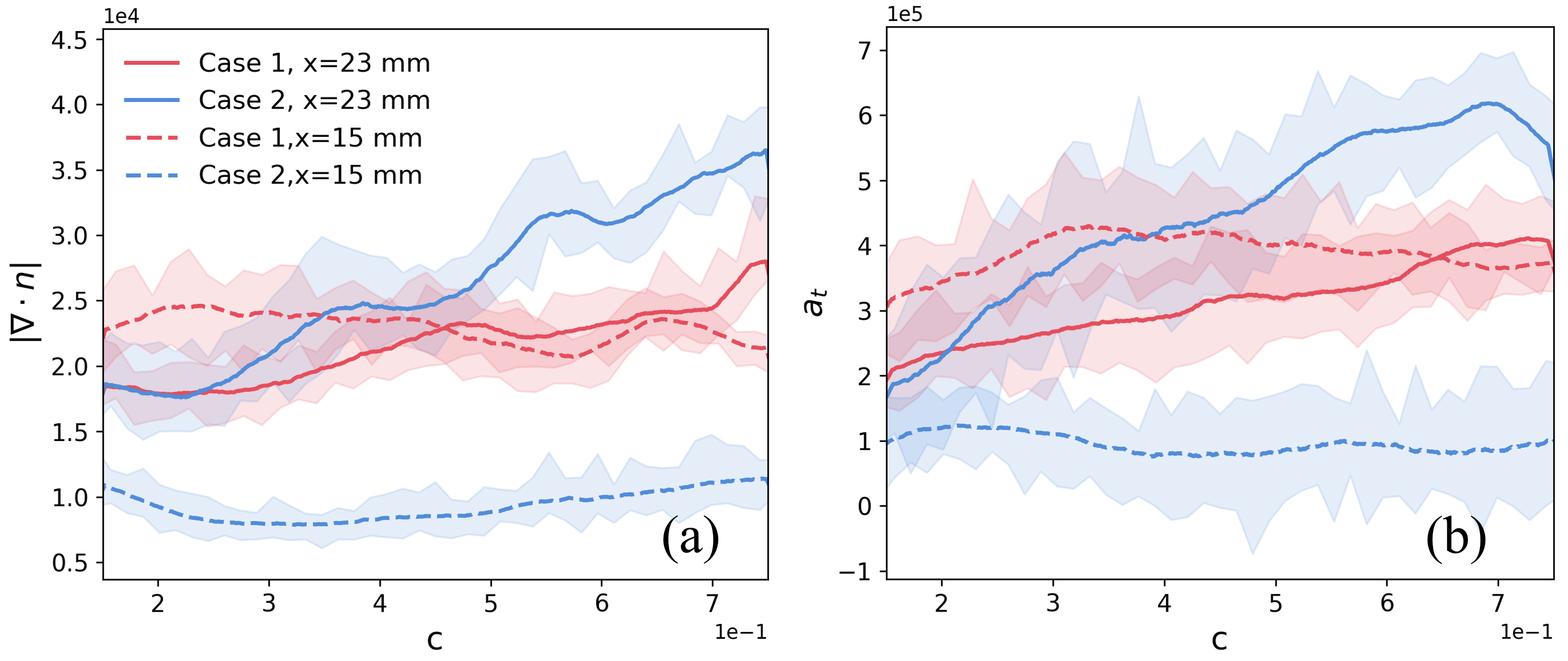}
    \fi
    \caption{(a) $|\nabla \cdot n|$ and (b) $a_t$ conditionally averaged on progress variable c. Colored areas depict distribution range of conditional means and lines represent smoothed results.} 
    \label{fig16:at_cur_c}
\end{figure} 
Figure.~\ref{fig16:at_cur_c} (a) and (b) depict the conditional averaged values of tangential strain rate $a_t$ and curvature magnitude $|\nabla \cdot n|$ as a function of progress variable across distinct regions. The comparison of Case 1 and Case 2 reveals that the $a_t$ and $|\nabla \cdot n|$ are notably elevated in Case 1 within the region approximately at $x=15$~mm. Combining the results of Case 1 depicted in Fig.~\ref{fig15:gradC_c}, the increase in averaged $a_t$ lead to the growth of $|\nabla c|$ within region where $c<0.3$; then with slight increase in $|\nabla \cdot n|$ and decrease in $a_t$ where $c>0.6$, a corresponding decrease is observed in $|\nabla c|$. Therefore, the collaborative interaction of the tangential strain rate and curvature serves to increase the flame thickness with better developed cavity shear layer. In addition, the prominent low curvature magnitude observed in Case 2 near $x=15$~mm play a vital role in facilitating the formation of a thinner flame. 

Regarding the analysis of the flame near $x=23$~mm, notable increases are observed in both $a_t$ and $|\nabla \cdot n|$ within Case 2. Also in both cases, $a_t$ and $|\nabla \cdot n|$ exhibits positive correlation with the growth of progress variable c. Given the increase of $|\nabla c|$ within regions where $c<0.35$ is apparent in Fig.~\ref{fig15:gradC_c} (b), it can be inferred that the heightened tangential strain rate primarily governs the variations in $|\nabla c|$ within this particular zone. As $a_t$ increases notably in Case 2 near $x=23$~mm, a reduction in flame thickness is observed. Following the rapid growth of curvature magnitude depicted in Fig.~\ref{fig16:at_cur_c} (a), the dissipation impact of curvature initiates a decline in $|\nabla c|$, thereby resulting in an increase of flame thickness within the oxidation zone. It is noteworthy to observe that this phenomenon is similar to which around $x=15$~mm, emphasizing the primary role of tangential strain rate in compressing the flame within the preheat layer, while the dissipation impact of curvature serves to increase flame thickness in the oxidation layer.

\section{Conclusions}
\label{sec:con}
High-resolution numerical simulations of premixed hydrogen/air flame over a cavity have been performed to understand the flame stabilization and structure under a supersonic inflow condition. Two cases with turbulent and laminar inlet conditions have been carried out to learn the effect of inflow boundary-layer turbulence. 

After the combustion reached the general steady-state feature, the overview of the flow and flame structure was first analysed. Combustion occurred within the cavity shear layer in both cases and propagated downstream along the lower wall. Different from cold condition, a weak compression wave was formed at the separation corner. Because of combustion in lower boundary layer, the jet flow was compressed and first reflected wave generally moved forward. After calculated time-averaged wall pressure difference along the lower wall, larger resistance was observed in the laminar case due to the intensified collision. With turbulent inflow, the fully developed cavity shear layer slightly depressed into the cavity and led to gentler collision at the aft wall, thereby resulting in diminished pressure drag.

To analysis flame stabilisation, studies on gas exchange and transport processes were carried out. Analysis of the mean flow streamlines indicated that the primary vortex in Case 1 exhibited a slightly greater size and depth. Then the introduction of marker and time-averaged mass flow rate demonstrated a pronounced mass exchange process near the rear wall in Case 2, whereas the wall-bounded turbulence facilitated a more uniformly distributed mass exchange process in Case 1. Additionally, the entered gas exhibited significant accumulation in the back part of the cavity in Case 2, indicating a weaker interaction between the primary and secondary vortices. This phenomenon suggested an extended residence time in the forepart of the cavity with laminar inflow, thereby promoting more complete reaction processes.

After that, flame structures under different inflow conditions were compared. In this section, reaction progress variable was defined and studied. Distribution of progress variable indicated that vortices were predominantly distributed within the expansive oxidation zone, and the thickness of the oxidation layer in Case 2 increased significantly due to the impingement of the flame on the rear wall. To further analysis the flame and turbulence interaction, flame stretch was investigated via tangential strain rate and flame surface curvature. The observed phenomena in Case 1 were similar to those reported in prior direct numerical simulations of premixed jet flames, whereas the behaviours in front part of cavity shear layer were differ due to the presence of roll-up vortices and the vigorous impingement on the rear wall in Case 2. As for the flame thickness, the presence of roll-up vortex resulted in a reduced flame thickness within regions exhibiting large positive curvature. With dissipation of this specific structure, the variation of flame thickness between these two cases became similar. Overall, the influence of tangential strain rate and curvature was consistent with the preceding results and the evolution of flame thickness was caused by the combined effect of the two factors.

Performances analysed in this study pave the way to design and optimization for the wall cavity as flame-holding approach. The application of premixed reactants could simulate the combustion of fully mixed condition, but also ignores the effect of injection. Therefore the combustion performances of different fuel injection conditions still need to be further investigated and discussed with high resolution.  


\section*{Acknowledgements}
This work is supported by the National Science Foundation of China (Grant Nos. 92270203 and 52276096).
J.F. also gratefully acknowledges the support of UK Engineering and Physical Sciences Research Council (EPSRC) through the Computational Science Centre for Research Communities (CoSeC) and the UK Turbulence Consortium (Grant No. EP/X035484/1). This work used the ARCHER2 UK National Supercomputing Service (https://www.archer2.ac.uk).
Part of the numerical simulations was also performed on the High Performance Computing Platform of CAPT of Peking University. The support from the Royal Society (IEC\textbackslash NSFC\textbackslash 223421) is also acknowledged.





\bibliography{main}

\end{document}